\documentclass[a4paper,11pt]{article}
\usepackage{jheppub} % for details on the use of the package, please
\usepackage{graphicx,wrapfig}

\usepackage{amsmath}

\usepackage{amssymb}
\usepackage{amsfonts}
\usepackage{amsbsy}
\usepackage{url}
\usepackage{enumerate}
\usepackage{caption}
\usepackage{color}
\usepackage{ulem}
\usepackage{bm}
\usepackage{multirow,bigdelim}

\newcommand{\be}{\begin{eqnarray}}
\newcommand{\ee}{\end{eqnarray}}
\newcommand{\p}{\partial}

\makeatletter
\@addtoreset{equation}{section}
\makeatother

\newcommand\rsout{\bgroup\markoverwith{\textcolor{red}{\rule[0.5ex]{2pt}{0.4pt}}}\ULon}

\title{
Exhausting all exact solutions of BPS domain wall networks 
in arbitrary dimensions
}

\author[a,b]{Minoru Eto,}
\author[a]{Masaki Kawaguchi,}
\author[b,c]{Muneto Nitta,}
\author[a]{Ryotaro Sasaki}

\affiliation[a]{Department of Physics, Yamagata University,\\
Kojirakawa-machi 1-4-12, Yamagata, Yamagata 990-8560, Japan}
\affiliation[b]{Research and Education Center for Natural
Sciences, Keio University, Hiyoshi 4-1-1, Yokohama, Kanagawa 223-8521, Japan}
\affiliation[c]{
Department of Physics, Keio University, Hiyoshi 4-1-1, Yokohama, Kanagawa 223-8521, Japan}

\emailAdd{meto(at)sci.kj.yamagata-u.ac.jp}
\emailAdd{ddwbb.daigaku(at)gmail.com}
\emailAdd{nitta(at)phys-h.keio.ac.jp}

\abstract{
We obtain full moduli parameters for
generic non-planar BPS networks of domain walls 
in an extended Abelian-Higgs model with $N $ complex scalar fields, 
and exhaust all exact solutions
in the corresponding $\mathbb{C}P^{N -1}$ model. 
We develop a convenient description by grid diagrams which are polytopes 
determined by mass parameters of the model. 
To illustrate the validity of our method,
we work out non-planar domain wall networks 
for lower $N $ in $3+1$ dimensions. 
In general, the networks can have compact vacuum bubbles, 
which are finite vacuum regions surrounded by domain walls,  
when the polytopes of the grid diagrams have inner vertices, 
and the size of bubbles can be controlled  by 
moduli parameters. 
We also construct domain wall networks with 
bubbles in the shapes of 
the Platonic, Archimedean, Catalan, and Kepler-Poinsot solids. 
}
\preprint{YGHP-20-03}

\begin{document}
\maketitle

%%%%%%%%%%%%%%%%% I N T R O D U C T I O N %%%%%%%%%%%%%%%%%%

\section{Introduction}
\label{sec:intro}

It sometimes happens that systems have 
multiple discrete vacua or ground states, 
which is inevitable when a discrete symmetry is spontaneously broken. 
In such a case, there appear 
domain walls (or kinks) in general  
\cite{Manton:2004tk,Vachaspati:2006zz,Rajaraman:1982is} 
which are inevitably created 
during second order phase transitions 
\cite{Kibble:1976sj,Kibble:1980mv,Zurek:1985qw,Zurek:1996sj}. 
They are the simplest topological solitons 
appearing in various condensed matter systems 
such as magnets \cite{magnetism}, 
graphenes \cite{graphene}, 
carbon nanotubes, 
superconductors \cite{chiral-p-wave}, 
atomic Bose-Einstein condensates 
\cite{Takeuchi:2012ee},
and helium superfluids~\cite{volovik,Zurek:1985qw,Zurek:1996sj}, 
as well as  
high density nuclear matter 
\cite{Eto:2012qd,Yasui:2019vci}, 
quark matter 
\cite{Eto:2013hoa} 
and early Universe~\cite{Vilenkin:2000jqa,Kibble:1980mv}. 
In cosmology, 
cosmological domain wall networks are suggested as a candidate of dark matter and/or dark energy 
\cite{cosmology}.

As the cases of other topological solitons,  
domain walls can become  
Bogomol'nyi-Prasad-Sommerfield (BPS) states 
\cite{Bogomolny:1975de,Prasad:1975kr},  
 attaining the minimum energy for a fixed boundary condition
and satisfy first order differential equations
called BPS equations.
In such cases, one can often embed the theories to 
supersymmetric (SUSY) theories by appropriately adding fermion superpartners,
in which BPS solitons preserve some fractions of SUSY.
Their topological charges are central (or tensorial) 
charges of corresponding SUSY algebras.
The BPS domain walls in $3+1$ dimensions 
were studied extensively in field theories with both ${\cal N}=1$ SUSY \cite{Dvali:1996xe,Dvali:1996bg,Kovner:1997ca,Smilga:1997pf,Chibisov:1997rc,Kaplunovsky:1998vt,deCarlos:1999xk,Dvali:1999pk,Edelstein:1997ej,Naganuma:2000gu,Arai:2009jd,Arai:2011gg,Nitta:2014pwa,Gudnason:2016frn,Lee:2017kaj, Arai:2018tkf} 
and ${\cal N}=2$ SUSY 
\cite{Abraham:1992vb,Abraham:1992qv,Gauntlett:2000bd,Gauntlett:2000ib,Tong:2002hi,Lee:2002gv,Arai:2002xa,Arai:2003es,Losev:2003gs,Isozumi:2003rp,Shifman:2003uh,Isozumi:2004jc,Isozumi:2004va,Eto:2004vy,Eto:2005wf, Eto:2005cc,Eto:2006mz,Eto:2006uw,Eto:2008dm,
Isozumi:2004vg,Eto:2008mf}, 
see Refs.~\cite{Tong:2005un,Eto:2006pg,Shifman:2007ce,Shifman:2009zz} as a review.
They preserve a half of SUSY and thereby are called $\frac{1}{2}$ BPS states,
accompanied by SUSY central (tensorial) charges $Z_m$ 
$(m=1,2$; labeling spatial coordinates $x^m$ perpendicular to the domain wall)
as domain wall topological charges \cite{deAzcarraga:1989mza,Kovner:1997ca,Chibisov:1997rc}. 
In general, if several domain walls meet along a line, it forms a planar domain wall junction. 
In SUSY models, the planar domain wall junctions preserve a quarter SUSY \cite{Gibbons:1999np,Carroll:1999wr,Gorsky:1999hk},  therefore are called $\frac{1}{4}$ BPS states, 
accompanied by a junction topological charge $Y$ 
in addition to $Z_m$ ($m =1,2$).
The $\frac{1}{4}$ BPS domain wall junctions 
have been studied in theories with ${\cal N}=1$ SUSY  
\cite{Abraham:1990nz,Gabadadze:1999pp,Oda:1999az,Shifman:1999ri,Ito:2000zf,Binosi:1999vb,Nam:2000qe,Carroll:1999mu,Naganuma:2001br,Nitta:2014pwa} 
and ${\cal N}=2$ SUSY 
\cite{Kakimoto:2003zu,Eto:2005cp,Eto:2005fm,Eto:2005mx,Eto:2006bb,Eto:2007uc,Fujimori:2008ee,Shin:2018chr,Shin:2019buk,Kim:2020obf}.
In the ${\cal N}=2$ SUSY gauge models
not only planar domain wall junctions, but also 
planar domain wall {\it networks} as 
$\frac{1}{4}$ BPS states were constructed 
\cite{Eto:2005cp,Eto:2005fm}.
The low-energy effective action for normalizable modes 
within the networks
was obtained \cite{Eto:2006bb} 
and applied  to study of low-energy dynamics \cite{Eto:2007uc}.
Non-BPS planer domain wall networks were also studied in Refs.~\cite{Saffin:1999au,Bazeia:1999xi}.

Recently, the present authors proposed a model, 
a $D+1$ dimensional $U(1)$ gauge theory \cite{Eto:2020vjm}
admitting novel analytic solutions of the BPS
{\it single non-planar} domain wall junctions.
This model cannot be made supersymmetric 
but still admit stable BPS states   
so that we can use the well-known Bogomol'nyi completion technique
to derive BPS equations.
The model consists of $N $ charged complex scalar fields and $N' $ neutral scalar fields coupled to the $U(1)$ gauge field. In Ref.~\cite{Eto:2020vjm},
we restricted ourself to the special numbers $N  - 1 = N'  = D$ 
and imposed the invariance under the symmetric group ${\cal S}_{D+1}$ of the rank $D+1$,
which are the symmetry groups of the regular $D$-simplex.

In this paper, we investigate generic 
{\it non-planar networks} of BPS domain walls in $D+1$ dimensions. 
We consider the generic case of 
$N  \ge D+1$ imposing no discrete symmetry,
and exhaust all exact solutions with full moduli  of 
generic BPS non-planar networks of  domain walls 
in the infinite $U(1)$ gauge coupling limit
in which the model reduces to the $\mathbb{C}P^{N -1}$ model.
These are the first exact solutions 
of non-planar domain wall networks in $D$ dimensions ($D\ge3$).
We firstly derive the   BPS equations for generic 
Abelian gauge theories in $D+1$ dimensions.
Then, we partially solve them by the moduli matrix formalism
\cite{Eto:2006pg} 
and find all moduli parameters of the generic domain wall network solutions.
We then demonstrate several concrete non-planar networks in the $\mathbb{C}P^{N -1}$
model in $D=3$ for $N =4,5,6$.
In the case of $N  = 4$, the solution has only one junction at which four vacua meet. 
Network structures appear for $N  > 4$. In the case of $N  = 5$, we show two different types of networks exist in general. The first type has a vacuum bubble (a compact vacuum domain) 
surrounded by semi-infinite vacuum domains. Instead, the second type does not have any
bubbles but all the vacuum domains are semi-infinitely extended.
In the $N =6$ case, there are three different types according to the number of the vacuum bubbles, two, one or zero.
Finally, we find a connection to the well-known polyhedra known from ancient times. 
Indeed, we find the vacuum bubbles which are congruent with
the five Platonic solids. In addition, the Archimedean and the Catalan solids appear
as the vacuum bubbles. We also construct the Kepler-Poinsot star solids 
as domain wall networks.

This paper is organized as follows.
In Sec. \ref{sec:model} we introduce our model.
In Sec.~\ref{sec:BPS_eq}, we derive the   BPS equations and clarify the 
moduli space of the BPS solutions.
In Sec.~\ref{sec:CP}, we first consider the infinite $U(1)$ gauge coupling limit in which
the model reduces to the massive $\mathbb{C}P^{N -1}$ nonlinear sigma model.
We then exhaust all exact solutions with full moduli parameters for the   BPS equations. 
We further give several examples of non-planar networks in $D=3$.
In Sec.~\ref{sec:platon}, we study relations between the domain wall networks in $D=3$ and the classic solids, like the Platonic, Archimedean, Catalan, and Kepler-Poinsot solids.
Finally, we summarize our results and give a discussion in Sec.~\ref{sec:conclusion}.

%%%%%%%%%%%%%%%%% S E C T I O N 2 %%%%%%%%%%%%%%%%%%

\section{The model}
\label{sec:model}

We study a $U(1)$ gauge theory with
$N $ charged complex scalar fields $H^A$ ($A = 1,2,\cdots,N $) and $N' $ real scalar fields
$\Sigma^{A'}$ ($A'=1,2,\cdots,N' $) in $D+1$-dimensional spacetime.
The Lagrangian is given by
\be
{\cal L} &=& -\frac{1}{4e^2}F_{\mu\nu}F^{\mu\nu} 
+ \frac{1}{2e^2}\sum_{A'=1}^{N' }\partial_\mu \Sigma^{A'} \partial^\mu \Sigma^{A'}
+ D_\mu H (D^\mu H)^\dagger - V\label{eq:lag}\\
V &=&~ \frac{1}{2e^2}Y^2 
+ \sum_{A'=1}^{N' }\left(\Sigma^{A'}H - H M^{A'}\right) \left(\Sigma^{A'}H - H M^{A'}\right)^\dagger,
\label{eq:pot}
\ee
where $H$ is an $N $ component row vector made of $H^A$,
\be
H = \left(H^1,\ H^2,\ \cdots, H^{N }\right),
\ee
$Y$ is a scalar quantity defined by
\be
Y = e^2\left(v^2 - H H^\dagger\right),
\ee
and $M^{A'}$ ($A' =1,\cdots,N' $) are 
$N $ by $N $ real diagonal mass matrices defined by
\be
M^{A'} = {\rm diag}\left(m_{A',1},\ m_{A',2},\ \cdots,\ m_{A',N }\right).
\ee
The spacetime index $\mu$ runs from $0$ to $D$, and $F_{\mu\nu}$ is 
a $U(1)$ gauge field strength.
The coupling constants in the Lagrangian in Eq. (\ref{eq:lag}) are taken 
to be the so-called Bogomol'nyi limit.
For later use, let us define $\bm{m}_A$ by an $N' $ vector whose components are the $A$th diagonal elements of $M^{A'}$s, namely,
\be
\bm{m}_A = \left(m_{1,A},\ m_{2,A},\ \cdots,\ m_{N' ,A}\right).
\ee
In the following, we will mostly consider generic masses  
\be
\bm{m}_A \neq \bm{m}_B,\qquad \text{if } A\neq B.
\label{eq:gene_cond}
\ee

Since the scalar potential $V$ is positive semidefinite, a classical vacuum of the theory is determined by $V=0$:
$
HH^\dagger = v^2$, and $\Sigma^{A'} H - H M^{A'} = 0$.
In the generic case of Eq.~(\ref{eq:gene_cond}), there are $N $ discrete vacua.
The $A$th vacua which we will denote by $\left<A\right>$ is given by
$H^B = v \delta^B_A$, and $\Sigma^{B'} = m_{B',A}$.
Simply, the vacua can be identified to the discrete points determined by the mass vectors $\{\bm{m}_A\}$ in the $\Sigma$ space:
\be
\left<A\right> :\bm{\Sigma} = \bm{m}_A.
\label{eq:vac}
\ee

The Lagrangian (\ref{eq:lag}) is motivated by supersymmetry.
In fact, when $N_{\rm F}'=2$, it is a bosonic part of 
an ${\cal N}=2$ supersymmetric theory 
with eight supercharges in four dimensions.

%%%%%%%%%%%%%%%%% S E C T I O N 3 %%%%%%%%%%%%%%%%%%

\section{Solving BPS equations for domain wall networks}
\label{sec:BPS_eq}

\subsection{Derivation of the BPS equations}

From now on, we investigate BPS states of ${\cal L}$ in Eq.~(\ref{eq:lag}) in the case that the number $N' $ of flavors for real adjoint scalars 
is equal to the number of the spatial dimensions $D$.\footnote{
In Ref.~\cite{Eto:2020vjm},
we considered the same model with restricted to a special case in which 
the %other 
flavor number $N $ is also related to the spatial dimensions, 
$N  = D+1$.
Instead, this work studies the case with generic $N  (\ge D+1)$.  }
In what follows, the Roman index $m$ stands not only for the spacial index as $m=1,2,\cdots,D$
but also for the index of $N' $ ($m\equiv A'$).
Then, the standard Bogomol'nyi completion for this system goes as follows:
\be
{\cal E} &=& \frac{1}{2e^2}\sum_{m>n}\left\{
F_{mn}^2 + \left(\partial_m \Sigma_n - \xi_m\xi_n \partial_n \Sigma_m\right)^2 \right\}
+ \frac{1}{2e^2}  \left(\sum_m\xi_m\partial_m\Sigma_m - Y\right)^2
\nonumber\\
&\ &+~ \sum_m \left\{D_m H + \xi_m(\Sigma_m H - H M_m)\right\} \left\{D_m H + \xi_m(\Sigma_m H - H M_m)\right\}^\dagger
\nonumber\\
&\ &+~
\sum_m  \xi_m {\cal Z}_m + \sum_{m>n}\xi_m\xi_n {\cal Y}_{mn}+ \sum_m \partial_m {\cal J}_m,
\label{eq:Bog}
\ee
with $\xi_m = \pm 1$, and 
we have defined the domain wall topological charge density ${\cal Z}_m$, 
the domain wall junction charge density
${\cal Y}_{mn}$, and ${\cal J}_m$ by
\be
{\cal Z}_m &=& v^2\partial_m \Sigma_m, \label{eq:Z_density}\\
{\cal Y}_{mn} &=& 
- \frac{1}{e^2}\det\left(
\begin{array}{cc}
\p_m \Sigma_m & \p_m\Sigma_n\\
\p_n \Sigma_m & \p_n \Sigma_n
\end{array}
\right),
\label{eq:Y_density}\\
{\cal J}_m &=& -\xi_m\left(\Sigma_m H - H M_m\right)H^\dagger,
\ee
respectively. 
The contribution by ${\cal J}_m$ vanishes under the space integrations 
since it is asymptotically zero
because of the vacuum condition $\Sigma_m H - H M_m = 0$.

The domain wall tension $Z_m$ measured along the $x^m$ direction 
can be defined from ${\cal Z}_m$ by
\be
Z_m = \int^\infty_{-\infty}dx^m\ \xi_m{\cal Z}_m 
= v^2 \xi_m \bigg(\Sigma_m\big|_{x^m=+\infty} - \Sigma_m\big|_{x^m=-\infty}\bigg) \ge 0,\  (\mbox{no sum over } m).
\label{eq:DW_tension}
\ee
Note that $Z_m$ is always {\it positive} regardless of
the choice of $\xi_m$. Hence, the genuine tension 
measured along the normal direction to the domain wall 
interpolating the vacua $\left<A\right>$
and $\left<B\right>$ is given by
\be
\left|\bm{Z}\right| = v^2 \left|\bm{m}_A - \bm{m}_B\right|.
\ee

On the other hand, 
the domain wall junction charge
can be defined from the topological charge density ${\cal Y}_{mn}$ as
\be
Y_{mn} =  \xi_m\xi_n \int dx^mdx^n\,{\cal Y}_{mn} = - \frac{\xi_m\xi_n}{e^2} S_{mn} \le 0,
\ee
negatively contributing to the BPS energy.
Here, we have defined
\be
S_{mn} \equiv \int dx^mdx^n\, \det\left(
\begin{array}{cc}
\p_m \Sigma_m & \p_m\Sigma_n\\
\p_n \Sigma_m & \p_n \Sigma_n
\end{array}
\right).
\ee
The integrand is 
a Jacobian of a map from the whole $x^m$-$x^n$ plane 
to a region in the $\Sigma^m$-$\Sigma^n$ plane defined 
by the function $(\Sigma_m(x^m,x^n),\Sigma_n(x^m,x^n))$ with all other coordinates
$x^k$ ($k\neq m,n$) fixed.
$S_{mn}$ can be either positive or negative, and its absolute value is the area of the image.
Nevertheless $Y_{mn}$ is always {\it negative} 
since $S_{mn}$ is accompanied with $(-\xi_m\xi_n)$,
and so it should be understood as a sort of binding energy among domain walls \cite{Oda:1999az,Shifman:1999ri,Ito:2000zf}.

Since the first three terms of Eq.~(\ref{eq:Bog}) are positive semidefinite, the Bogomol'nyi energy bound is given by
\be
{\cal E} \ge
\sum_m  \xi_m {\cal Z}_m + \sum_{m>n}\xi_m\xi_n {\cal Y}_{mn}+ \sum_m \partial_m {\cal J}_m,
\ee
and it is saturated by the BPS states satisfying the BPS equations
\be
F_{mn} = 0,\label{eq:BPS1}\\
\xi_n\partial_m \Sigma_n - \xi_m \partial_n \Sigma_m = 0,\label{eq:BPS2}\\
\xi_mD_m H + \Sigma_m H - H M_m = 0,\label{eq:BPS3}\\
\sum_m\xi_m\partial_m\Sigma_m - Y = 0,\label{eq:BPS4}
\ee
where $m,n=1,2,\cdots,D$.
One can verify that all solutions of the above BPS equations solve the full equations of motion.
This is the $D$ dimensional extension of the $\frac{1}{4}$ BPS equations of the planar domain wall junction in $D=2$ cases
studied in Refs.~\cite{Kakimoto:2003zu,Eto:2005cp,Eto:2006bb,Eto:2007uc}.

In the previous work \cite{Eto:2020vjm} 
by the present authors, we generalized 
the topological charge densities ${\cal Z}_m$ and ${\cal Y}_{mn}$ 
under the observation that ${\cal Z}_m$ and ${\cal Y}_{mn}$ 
are nothing but one- and two-dimensional Jacobians of maps $x^m \to \Sigma_m$
and $(x^m,x^n) \to (\Sigma_m,\Sigma_n)$, respectively. 
For a domain wall interpolating the $\left<A\right>$ and $\left<B\right>$ vacua, 
an integral of ${\cal Z}_m$ 
(with an appropriate rescale to make it dimensionless) 
measures a covering number  
of a map from
$\mathbb{R}^1 (-\infty < x^m < \infty)$ onto the one-dimensional interval 
$(m_{m,A} < \Sigma_m < m_{m,B})$. The charge is topological and indeed takes values 
either $+1$, 0, or $-1$. Similar arguments hold for ${\cal Y}_{mn}$, and it is straightforward 
to generalize it in $D$ dimensions as
\be
{\cal W}_d(m_1,m_2,\cdots,m_d) = \det
\left(
\begin{array}{cccc}
\p_{m_1}\Sigma_{m_1} & \p_{m_1}\Sigma_{m_2} & \cdots & \p_{m_1}\Sigma_{m_d}\\
\p_{m_2}\Sigma_{m_1} & \p_{m_2}\Sigma_{m_2} & \cdots & \p_{m_2}\Sigma_{m_d}\\
\vdots & & \ddots & \vdots \\
\p_{m_d}\Sigma_{m_1} & \p_{m_d}\Sigma_{m_2} & \cdots & \p_{m_d}\Sigma_{m_d}
\end{array}
\right),
\label{eq:W_d}
\ee
where $1 \le d \le D$, $m_\alpha \in \{1,2,\cdots,D\}$ ($\alpha=1,2,\cdots,d$) 
and $m_\alpha > m_\beta$ if $\alpha > \beta$.
We have ${\cal W}_1(m) \propto {\cal Z}_m$ and ${\cal W}_{2}(m,n) \propto {\cal Y}_{mn}$.
In general, the BPS solutions in $D$ dimensions have $d$ dimensional substructures,
and ${\cal W}_d$ provides us topological numbers associated with the substructures.

\subsection{The moduli matrix method}

Let us solve the BPS equations (\ref{eq:BPS1})--(\ref{eq:BPS4}).
To this end, we first introduce a complex scalar function $S(x^m)$ by
\be
A_m - i \xi_m \Sigma_m = - i\p_m \log S.
\label{eq:S}
\ee
Then, Eqs.~(\ref{eq:BPS1}) and (\ref{eq:BPS2}) are automatically satisfied.
Plugging this into Eq.~(\ref{eq:BPS3}), we have
\be
\p_m H + (\p_m\log S) H - \xi_m H M_m = 0.
\ee
This can be solved by
\be
H = v S^{-1}H_0e^{\xi_m M_m x^m},
\label{eq:H}
\ee
where $H_0$ is an arbitrary complex constant $N $ vector. 
$H_0$ is called the moduli matrix
because the elements of $H_0$ are moduli (integration constants) of the BPS solutions.
Finally, we are left with the fourth equation (\ref{eq:BPS4}). To solve this, let us express
the neutral real scalar fields $\Sigma_m$ in terms of $S$:
\be
\Sigma_m = \frac{1}{2}\xi_m\p_m \log \Omega,\qquad \Omega \equiv |S|^2,
\label{eq:Sigma}
\ee
where $\Omega$ is the gauge invariant quantity. Using this, the fourth BPS equation (\ref{eq:BPS4}) 
is cast into the following Poisson equation in $D$ dimensions
\be
\frac{1}{2}\nabla^2 \log \Omega = e^2v^2\left(1 - \Omega^{-1} H_0e^{2\xi_m M_m x^m} H_0^\dagger\right).
\label{eq:master}
\ee
We call this the master equation for the   BPS states.

We note that the original fields $A_m$, $\Sigma_m$,
and $H$ are intact under the following transformation
\be
(S,H_0) \to V(S,H_0),\qquad V \in \mathbb{C}^*,\label{eq:V-transf}
\ee
where $V$ is constant. This is called the $V$-transformation under which
any physical informations are independent.

To solve the master equation, we first need to fix the moduli matrix $H_0$.
Once $H_0$ is given, the boundary condition at $|\bm{x}| \to \infty$ is automatically specified.
Namely, we should solve the master equation with the boundary condition
\be
\lim_{|\bm{x}| \to \infty} \Omega = H_0e^{2\xi_m M_m x^m} H_0^\dagger.
\ee

As a trivial example, let us consider a homogeneous vacuum, say the first vacuum $\left<1\right>$.
The $\left<1\right>$ vacuum configuration is given by the moduli matrix
\be
H_0 = \left(h_1,\ 0,\ \cdots,\ 0\right),\qquad h_1 \in \mathbb{C}^*.
\label{eq:MM_vac1}
\ee
The corresponding master equation becomes
\be
\frac{1}{2}\nabla^2 \log \Omega = e^2v^2\left(1 - \Omega^{-1} |h_1|^2 e^{2\xi_m m_{m,1}x^m} \right),
\ee
which can be solved by
\be
\Omega = |h_1|^2e^{2\xi_m m_{m,1}x^m}.
\ee
Plugging this into Eq.~(\ref{eq:Sigma}), we immediately find $\bm{\Sigma} = \bm{m}_1$.
It is also straightforward to verify $H = (v,\ 0,\ \cdots,\ 0)$.
Note that the constant $h_1$ in the moduli matrix does not play any role in the above example.
This redundancy comes from 
the $V$-transformation in Eq.~(\ref{eq:V-transf})
Indeed, we could, from the first point, fix
the moduli matrix (\ref{eq:MM_vac1}) as $H_0 = \left(h_1,\ 0,\ \cdots,\ 0\right) \to (1,\ 0,\ \cdots,\ 0)$.

Nontrivial inhomogeneous solutions including a single domain wall connecting the $\left<A\right>$ and $\left<B\right>$ vacua
are generated by the moduli matrix with the non-zero constants only in the $A$th and $B$th entries as 
\be
H_0 
&=& \left(0,\ \cdots,\ 0,\ 1,\ 0,\ \cdots,\ 0,\ h_B,\ 0,\ \cdots,\ 0\right)
\nonumber\\
&\sim&
\left(0,\ \cdots,\ 0,\ h_A,\ 0,\ \cdots,\ 0,\ 1,\ 0,\ \cdots,\ 0\right).
\label{eq:MM_DW}
\ee
The non-zero moduli parameters $h_A$ and $h_B$ are related by the $V$-transformation as $h_Ah_B = 1$.

Similarly, if we have $H_0$ with three non-zero constants, we will have a domain wall junction dividing the corresponding 
three vacua. The moduli matrix is given by
\be
H_0 
= \left(0,\ \cdots,\ 0,\ 1,\ 0,\ \cdots,\ 0,\ h_B,\ 0,\ \cdots,\ 0,\ h_C,\ 0,\ \cdots,\ 0\right).
\ee

Maximally complex solutions dividing $N $ vacuum domains in $D$ dimensions 
are, then, obtained by the moduli matrix which has no zeros in any elements.
Such complicated extended objects in $D$ dimensions are not easy for us to handle without exact solutions.
Unfortunately, the master equation (\ref{eq:master}) does not seem analytically solvable, except for the
very special cases possessing the highest discrete symmetry group ${\cal S}_{D+1}$, the symmetric group of the degree $D+1$,
in the model with the special number of the flavor $N  = D+1$ and finely tuned parameters $g$, $c$, and $m_{A',A}$, 
as demonstrated in Ref.~\cite{Eto:2020vjm}.
Even in the fine tuned models, the exact analytic solution were only found for a single junction of the domain walls since $N =D+1$ is the minimum number.
However, as we show in the next section, the solutions include 
domain wall networks when $N  > D+1$.

To see shapes of domain wall networks, 
let us define a weight for each vacuum by
\be
w^{\left<A\right>} = e^{\tilde{\bm m}_A \cdot \bm{x} + a_A},\quad (A = 1,\cdots,N ).
\ee 
Since the weight is the exponential function of the spatial coordinate,
only one weight dominates the rest $N -1$ weights at each point $\bm{x}$.
Suppose the weight of the $\left<A\right>$ vacuum is dominant in vicinity of a point $\bm{x}_0$.
There, we have
$\Omega \sim (w^{\left<A\right>})^2$,
and then $\bm{\Sigma}$ reads from Eq.~(\ref{eq:Sigma})
\be
\bm{\Sigma}\big|_{\bm{x} \sim \bm{x}_0} = \frac{1}{2}\tilde\nabla \log (w^{\left<A\right>})^2 = \bm{m}_A.
\ee
This implies that the region where the weight $w^{\left<A\right>}$ is dominant corresponds to
the vacuum $\left<A\right>$. 

Let us next consider a situation that the vacua $\left<A\right>$ and $\left<B\right>$ are
next to each other. We can estimate where the two vacua transit by comparing the weights of
those vacua. The transition occurs at points on which the two weights are equal.
This condition determines a hyperplane which is a subspace of codimension 1 in $\mathbb{R}^D$:
\be
\left<A,B\right>\ :\quad
w^{\left<A\right>} = w^{\left<B\right>} \quad \Leftrightarrow \quad
\left(\tilde{\bm{m}}_A-\tilde{\bm{m}}_B\right)\cdot\bm{x} + a_A-a_B = 0.
\label{eq:hyperplane}
\ee
This is a straightforward generalization of $D=2$ \cite{Eto:2005cp} to generic $D$ dimensions. 
This hyperplane is nothing but a domain wall interpolating the vacua $\left<A\right>$ and 
$\left<B\right>$,
and we call it a 1-wall.

The three vacua, say $\left<A\right>$, $\left<B\right>$ and $\left<C\right>$,
can happen to be adjacent at a hyperplane of codimension 2, which is conventionally called
the domain wall junction. We call it a 2-wall. The position of the 2-wall
corresponds to the region where the thee weights are equal as
\be
\left<A,B,C\right>\ :\quad w^{\left<A\right>} = w^{\left<B\right>} = w^{\left<C\right>}.
\ee

These can be naturally generalized to a $d$-wall which is a $d$ codimensional intersection
dividing $d+1$ vacua. The position of the $d$-wall is defined by
\be
\left<A_1,A_2,\cdots,A_{d+1}\right>\ :\quad
w^{\left<A_1\right>} = w^{\left<A_2\right>} = \cdots = w^{\left<A_{d+1}\right>}.
\ee
In the next section, we will see that the position of the $d$-wall estimated by the weight
is related to the generalized topological charge ${\cal W}_d$ defined in Eq.~(\ref{eq:W_d}).

%%%%%%%%%%%%%%%%% S E C T I O N 4 %%%%%%%%%%%%%%%%%%

\section{Exhausting all exact solutions of domain wall networks in the $\mathbb{C}P^{N -1}$ model}
\label{sec:CP}

\subsection{General solutions}
There is a great simplification allowing us to obtain all exact solutions for generic domain wall
networks in $D$ dimensions. It is the infinite gauge coupling limit in which we formally send the gauge coupling $e$
to infinity in the Lagrangian (\ref{eq:lag}). There, the kinetic terms of $A_\mu$ and $\Sigma_m$ vanish to become Lagrange multipliers. At the same time, the first term in the potential (\ref{eq:pot})
forces the charged fields $H^A$ to take their values in the restricted region $S^{2N -1}$ defined by
$HH^\dagger = v^2$. Furthermore, the overall phase of $H_A$ is gauged. Therefore, the physical target space in
the infinite gauge coupling limit is reduced to 
the complex projective space
\be
\mathbb{C}P^{N -1} \simeq \frac{SU(N )}{SU(N -1)\times U(1)} \simeq \frac{S^{2N -1}}{S^1}.
\ee
Indeed, if we eliminate the gauge field $A_\mu$ from ${\cal L}\big|_{e\to\infty}$, it reduces to the standard
Lagrangian of the nonlinear $\mathbb{C}P^{N -1}$ model. 
Similarly, if we eliminate the
neutral scalar fields $\Sigma_m$, we get a non-trivial potential which lifts all the points  of the 
$\mathbb{C}P^{N -1}$ target space leaving 
$N $ discrete points as vacua.

Here, we do not eliminate the auxiliary fields $A_\mu$ and $\Sigma_m$. 
The   BPS equations in 
Eqs.~(\ref{eq:BPS1})--(\ref{eq:BPS4}) remain the same, 
and the first three equations 
(\ref{eq:BPS1})--(\ref{eq:BPS3}) are solved by Eqs.~(\ref{eq:S}) and (\ref{eq:H}).
The fourth equation (\ref{eq:BPS4}) rewritten as Eq.~(\ref{eq:master}) reduces 
to an algebraic equation, easily solved by
\be
\Omega\big|_{e\to\infty} = H_0e^{2\xi_m M_m x^m} H_0^\dagger.
\ee
Thus, we have completely solved the   BPS equations for arbitrary moduli matrix $H_0$ in the infinite gauge coupling limit.
We would like to emphasize that
this is the first solutions of the domain wall networks in $D\ge3$.

For later convenience, let us rewrite this in a more useful form. Firstly, let us denote $H_0$ as
\be
H_0 = \left(e^{a_1+ib_1},\ e^{a_2+ib_2},\ \cdots,\ e^{a_{N }+ib_{N }}\right),
\ee
where $\{a_A\}$ and $\{b_A\}$ are $N $ real parameters which we restrict to satisfy the conditions 
$\sum_A a_A = \sum_A b_A = 0$ by using the $V$-transformation. Furthermore, let us define the new mass vectors
\be
\tilde{\bm{m}}_A = \left(\xi_1 m_{1,A},\ \xi_2 m_{2,A},\ \cdots,\ \xi_{D} m_{D,A}\right).
\ee
Then, we have
\be
\Omega\big|_{e\to\infty} = \sum_{A=1}^{N } e^{2(\tilde{\bm{m}}_A\cdot \bm{x}+a_A)} .
\ee
By using this, the BPS energy density can be simply expressed by
\be
{\cal E}\big|_{e\to\infty} = v^2 \tilde \nabla \cdot \bm{\Sigma} 
= \frac{v^2}{2} \nabla^2 \log \left(\sum_{A=1}^{N } e^{2(\tilde{\bm{m}}_A\cdot \bm{x}+a_A)}\right),
\label{eq:ene_strong}
\ee
where we have defined
\be
\tilde\nabla = \left(\xi_1\p_1,\ \xi_2\p_2,\ \cdots,\ \xi_D \p_D\right),\qquad
\tilde{\nabla}^2 = \nabla^2,
\ee
and used $\bm{\Sigma} = \frac{1}{2}\tilde\nabla \log\Omega$.
From this expression, we can see that the $N -1$ real parameters $\{a_A\}$ are the moduli which
relate to the shape of the networks in the real space. On the other hand, the other parameters $\{b_A\}$ are internal moduli of $U(1)^{N -1}$ associated with constituent domain walls.

The domain walls for $D=1$ and the domain wall networks for $D=2$ have been studied very well
in the literature, therefore, we will concentrate on $D=3$ in the following subsections.

%%%%%%%%%%%%%%%%% S E C T I O N 5 %%%%%%%%%%%%%%%%%%

\subsection{$N =4$: Tetrahedron: Single domain wall junction}

\begin{figure}[ht]
\begin{center}
\includegraphics[height=5cm]{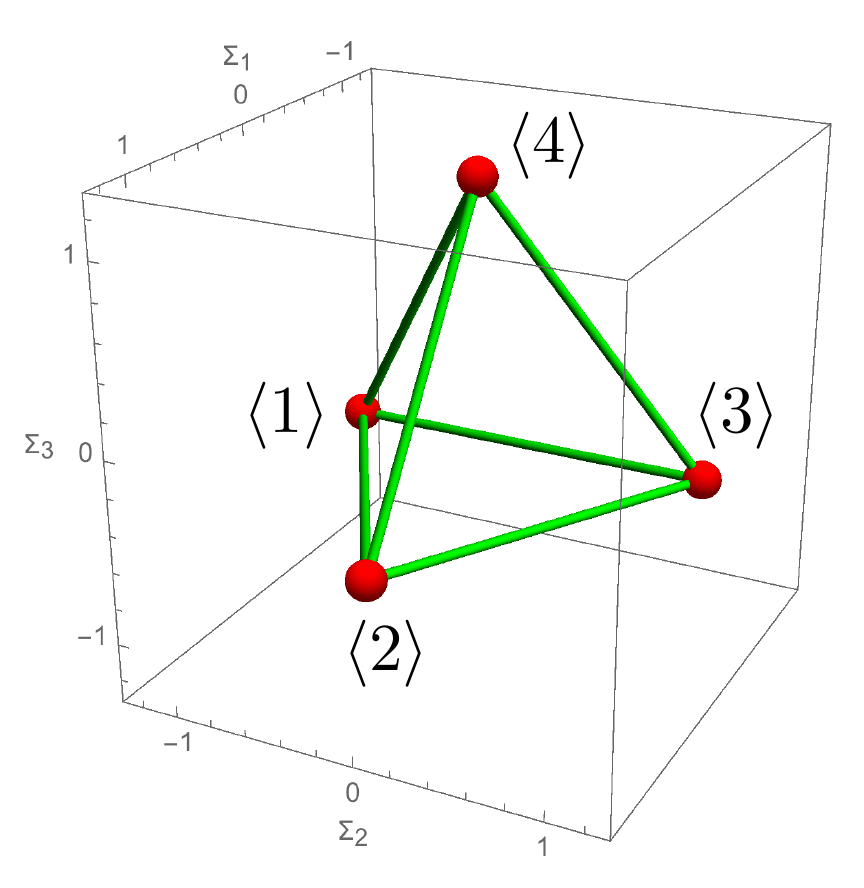}
\caption{The grid diagram ($\bm{\Sigma}$ space): 
The four vacua (red points) in $N  = 4$ model are shown.}
\label{fig:Nf_4_grid}
\end{center}
\end{figure}

Let us start with a model with $N =4$.
There are $N =4$ vacua which are the minimal numbers 
for a nonplanar domain wall junction to exist.
For simplicity, let us set the four mass vectors $\bm{m}_A$ to be four vertices of
a regular tetrahedron as
\be
\bm{m_1} &=& \left(-\frac{1}{\sqrt{3}},-1,-\frac{1}{\sqrt{6}}\right),\label{eq:M_tetra_1}\\
\bm{m_2} &=& \left(\frac{2}{\sqrt{3}},0,-\frac{1}{\sqrt{6}}\right),\\
\bm{m_3} &=& \left(-\frac{1}{\sqrt{3}},1,-\frac{1}{\sqrt{6}}\right),\\
\bm{m_4} &=& \left(0,0,\sqrt{\frac{3}{2}}\right).\label{eq:M_tetra_4}
\ee 
Then, the four vacua correspond to four vertices of the regular tetrahedron in
the $\Sigma_1$-$\Sigma_2$-$\Sigma_3$ space, as shown 
in Fig.~\ref{fig:Nf_4_grid}. 
We call polyhedrons drawn in the $\bm{\Sigma}$ space as
grid diagrams \cite{Eto:2005cp}.
Note that we have taken the regular tetrahedron (\ref{eq:M_tetra_1}) -- (\ref{eq:M_tetra_4}) just for simplicity.
In general, the grid diagrams do not have to be congruent 
with a regular tetrahedron. The following arguments are valid for the generic grid diagrams.

As we explained in Eq.~(\ref{eq:MM_DW}), single domain walls (1-walls) can be described by the 
moduli matrix with non-vanishing elements. For instance, 
the moduli matrix $H_0 = (e^{a+ib},e^{-a-ib},0,0)$ 
yield a domain wall $\left<1,2\right>$ 
connecting the $\left<1\right>$ and $\left<2\right>$ vacua.
The corresponding $\Omega$ is $\Omega = w^{\left<1\right>} + w^{\left<2\right>}$, and
the exact domain wall solution is given 
\be
\Sigma_1 
&=& \frac{m_{1,1} e^{2(\tilde{\bm m}_1\cdot  \bm{x}+a)} + m_{1,2}e^{2(\tilde{\bm m}_2\cdot  \bm{x}-a)}}{e^{2(\tilde{\bm m}_1\cdot  \bm{x}-a)} + e^{2(\tilde{\bm m}_2\cdot  \bm{x}-a)}},\\
\Sigma_2 
&=& \frac{m_{2,1} e^{2(\tilde{\bm m}_1\cdot  \bm{x}+a)} + m_{2,2}e^{2(\tilde{\bm m}_2\cdot  \bm{x}-a)}}{e^{2(\tilde{\bm m}_1\cdot  \bm{x}-a)} + e^{2(\tilde{\bm m}_2\cdot  \bm{x}-a)}},\\
\Sigma_3 
&=& \frac{m_{3,1} e^{2(\tilde{\bm m}_1\cdot  \bm{x}+a)} + m_{3,2}e^{2(\tilde{\bm m}_2\cdot  \bm{x}-a)}}{e^{2(\tilde{\bm m}_1\cdot  \bm{x}-a)} + e^{2(\tilde{\bm m}_2\cdot  \bm{x}-a)}}.
\ee
Note that $\Sigma_i$ takes its value in the finite interval $\Sigma_i \in [m_{i,1},m_{i,2}]$
($i=1,2$)
when we sweep the real space $\mathbb{R}^3$. Namely, $\bm{\Sigma}$ connects 
the two vertices $\bm{m}_1$ and
$\bm{m}_2$, as desired. Thus, the domain wall solution $\bm{\Sigma}(\bm{x})$ can be seen
as a function from the real space $\mathbb{R}^3$ to the $\bm{\Sigma}$ space, and
its image is the linear segment between $\bm{m}_1$ and $\bm{m_2}$,
\be
\frac{\Sigma_1 - m_{1,1}}{\Sigma_1 - m_{1,2}}
= \frac{\Sigma_2 - m_{2,1}}{\Sigma_2 - m_{2,2}}
= \frac{\Sigma_3 - m_{3,1}}{\Sigma_3 - m_{3,2}}.
\ee
The topological charges of 1-walls associated with Eq.~(\ref{eq:W_d}) are
\be
W_1^{(x)}(\bm{m}_1,\bm{m}_2) &=& \int^\infty_{-\infty} dx\ \p_1 \Sigma_1
= \int~d\Sigma_1
= |m_{1,1}-m_{1,2}|,\\
W_1^{(y)}(\bm{m}_1,\bm{m}_2) &=& \int^\infty_{-\infty} dy\ \p_2 \Sigma_2
= \int~d\Sigma_2
= |m_{2,1}-m_{2,2}|,\\
W_1^{(z)}(\bm{m}_1,\bm{m}_2) &=& \int^\infty_{-\infty} dz\ \p_3 \Sigma_3
= \int~d\Sigma_3
= |m_{3,1}-m_{3,2}|.
\ee
These are the lengths of segments of the edge connecting $\left<1\right>$ and $\left<2\right>$
projected onto the axes $\Sigma_{1,2,3}$.
It is straightforward to construct the other domain walls connecting 
arbitrary pair of $\left<A\right>$
and $\left<B\right>$. They correspond to the edges of the tetrahedron in
Fig.~\ref{fig:Nf_4_grid}.

A domain wall junction (2-wall) connecting three vacua, say $\left<1\right>$,
$\left<2\right>$, and $\left<3\right>$, can be also constructed very easily.
One mere needs to prepare the moduli matrix
$H_0 = \left(e^{a_1 + ib_1},e^{a_2 + ib_2},e^{a_3 + ib_3},0\right)$ with three nonzero elements.
The exact solution is given by
\be
\Sigma_1 
&=& \frac{m_{1,1} e^{2(\tilde{\bm m}_1\cdot  \bm{x}+a_1)} 
+ m_{1,2}e^{2(\tilde{\bm m}_2\cdot  \bm{x}+a_2)} 
+ m_{1,3}e^{2(\tilde{\bm m}_3\cdot  \bm{x}+a_3)}}{
e^{2(\tilde{\bm m}_1\cdot  \bm{x}+a_1)} 
+ e^{2(\tilde{\bm m}_2\cdot  \bm{x}+a_2)}
+ e^{2(\tilde{\bm m}_3\cdot  \bm{x}+a_3)}},\\
\Sigma_2 
&=& \frac{m_{2,1} e^{2(\tilde{\bm m}_1\cdot  \bm{x}+a_1)} 
+ m_{2,2}e^{2(\tilde{\bm m}_2\cdot  \bm{x}+a_2)} 
+ m_{2,3}e^{2(\tilde{\bm m}_3\cdot  \bm{x}+a_3)}}{
e^{2(\tilde{\bm m}_1\cdot  \bm{x}+a_1)} 
+ e^{2(\tilde{\bm m}_2\cdot  \bm{x}+a_2)}
+ e^{2(\tilde{\bm m}_3\cdot  \bm{x}+a_3)}},\\
\Sigma_3 
&=& \frac{m_{3,1} e^{2(\tilde{\bm m}_1\cdot  \bm{x}+a_1)} 
+ m_{3,2}e^{2(\tilde{\bm m}_2\cdot  \bm{x}+a_2)} 
+ m_{3,3}e^{2(\tilde{\bm m}_3\cdot  \bm{x}+a_3)}}{
e^{2(\tilde{\bm m}_1\cdot  \bm{x}+a_1)} 
+ e^{2(\tilde{\bm m}_2\cdot  \bm{x}+a_2)}
+ e^{2(\tilde{\bm m}_3\cdot  \bm{x}+a_3)}}.
\ee
These $\bm{\Sigma}$ satisfies the equation
\be
\left\{(\bm{m}_3 - \bm{m}_1) \times (\bm{m}_2-\bm{m}_1)\right\}\cdot(\bm{\Sigma} - \bm{m}_1) = \bm{0},
\ee
representing the 2 dimensional plane on which the three points $\bm{m}_1$, $\bm{m}_2$, and $\bm{m}_3$ are located.
The image of the map $\bm{\Sigma}(\bm{x})$ in this case is the triangle whose vertices are
$\left<1\right>$, $\left<2\right>$, and $\left<3\right>$.
The corresponding topological charges are identical to the areas of the projections
of the triangle onto the three planes ($\Sigma_1$--$\Sigma_2$, $\Sigma_2$--$\Sigma_3$,
and $\Sigma_3$--$\Sigma_1$)  as
\be
W_2^{(xy)}(\bm{m}_1,\bm{m}_2,\bm{m}_3) &=& \int^\infty_{-\infty} dxdy\ 
(\p_1\Sigma_2 \p_2\Sigma_1 - \p_1\Sigma_1\p_2\Sigma_2) \nonumber\\
&=& \int~d\Sigma_1d\Sigma_2 
= \frac{1}{2}\big|\left[(\bm{m}_1-\bm{m}_2)\times (\bm{m}_1-\bm{m}_3)\right]_3\big|,\\
W_2^{(yz)}(\bm{m}_1,\bm{m}_2,\bm{m}_3) &=& \int^\infty_{-\infty} dydz\ 
(\p_2\Sigma_3 \p_3\Sigma_2 - \p_2\Sigma_2\p_3\Sigma_3) \nonumber\\
&=& \int~d\Sigma_2d\Sigma_3 = \frac{1}{2}\big|\left[(\bm{m}_1-\bm{m}_2)\times (\bm{m}_1-\bm{m}_3)\right]_1\big|,\\
W_2^{(zx)}(\bm{m}_1,\bm{m}_2,\bm{m}_3) &=& \int^\infty_{-\infty} dzdx\ 
(\p_3\Sigma_1 \p_1\Sigma_3 - \p_3\Sigma_3\p_1\Sigma_1) \nonumber\\
&=& \int~d\Sigma_1d\Sigma_2 = \frac{1}{2}\big|\left[(\bm{m}_1-\bm{m}_2)\times (\bm{m}_1-\bm{m}_3)\right]_2\big|.
\ee
Again, it is straightforward to construct the other 2-walls dividing 
arbitrary set of three vacua. They correspond to the faces of the tetrahedron in
Fig.~\ref{fig:Nf_4_grid}.

Finally, we come to the 3-wall $\left<1,2,3,4\right>$ consisting of the four vacua,
six 1-walls, and four 2-walls. It is described by the full moduli matrix
$H_0 = (e^{a_1+ib_1},e^{a_2+ib_2},e^{a_3+ib_3},e^{a_4+ib_4})$, and
the exact solution is given by
\be
\Sigma_1 
&=& \frac{m_{1,1} e^{2(\tilde{\bm m}_1\cdot  \bm{x}+a_1)} 
+ m_{1,2}e^{2(\tilde{\bm m}_2\cdot  \bm{x}+a_2)} 
+ m_{1,3}e^{2(\tilde{\bm m}_3\cdot  \bm{x}+a_3)}
+ m_{1,4}e^{2(\tilde{\bm m}_4\cdot  \bm{x}+a_4)}
}{
e^{2(\tilde{\bm m}_1\cdot  \bm{x}+a_1)} 
+ e^{2(\tilde{\bm m}_2\cdot  \bm{x}+a_2)}
+ e^{2(\tilde{\bm m}_3\cdot  \bm{x}+a_3)}
+ e^{2(\tilde{\bm m}_4\cdot  \bm{x}+a_4)}},\\
\Sigma_2 
&=& \frac{m_{2,1} e^{2(\tilde{\bm m}_1\cdot  \bm{x}+a_1)} 
+ m_{2,2}e^{2(\tilde{\bm m}_2\cdot  \bm{x}+a_2)} 
+ m_{2,3}e^{2(\tilde{\bm m}_3\cdot  \bm{x}+a_3)}
+ m_{2,4}e^{2(\tilde{\bm m}_4\cdot  \bm{x}+a_4)}}{
e^{2(\tilde{\bm m}_1\cdot  \bm{x}+a_1)} 
+ e^{2(\tilde{\bm m}_2\cdot  \bm{x}+a_2)}
+ e^{2(\tilde{\bm m}_3\cdot  \bm{x}+a_3)}
+ e^{2(\tilde{\bm m}_4\cdot  \bm{x}+a_4)}},\\
\Sigma_3 
&=& \frac{m_{3,1} e^{2(\tilde{\bm m}_1\cdot  \bm{x}+a_1)} 
+ m_{3,2}e^{2(\tilde{\bm m}_2\cdot  \bm{x}+a_2)} 
+ m_{3,3}e^{2(\tilde{\bm m}_3\cdot  \bm{x}+a_3)}
+ m_{3,4}e^{2(\tilde{\bm m}_4\cdot  \bm{x}+a_4)}}{
e^{2(\tilde{\bm m}_1\cdot  \bm{x}+a_1)} 
+ e^{2(\tilde{\bm m}_2\cdot  \bm{x}+a_2)}
+ e^{2(\tilde{\bm m}_3\cdot  \bm{x}+a_3)}
+ e^{2(\tilde{\bm m}_4\cdot  \bm{x}+a_4)}}.
\ee
Note that we can set $a_1=a_2=a_3=a_4$ by using the three translational symmetries and
the $V$-transformation, without loss of generality.
Now, the solution $\bm{\Sigma}(\bm{x})$ maps the real three dimensional space $\mathbb{R}^3$
to the tetrahedron itself in the $\bm{\Sigma}$ space.
The corresponding topological charge reads
\be
W_3^{(xyz)}(\bm{m}_1,\bm{m}_2,\bm{m}_3,\bm{m}_4) 
&=& \int^\infty_{-\infty}dxdydz~
\det \left(
\begin{array}{ccc}
\p_1 \Sigma_1 & \p_1 \Sigma_2 & \p_1\Sigma_3\\
\p_2 \Sigma_1 & \p_2 \Sigma_2 & \p_2\Sigma_3\\
\p_3 \Sigma_1 & \p_3 \Sigma_2 & \p_3\Sigma_3
\end{array}
\right)
= \int d\Sigma_1d\Sigma_2 d\Sigma_3 \nonumber\\
&=& \frac{1}{6}\big|\big((\bm{m}_2-\bm{m}_1)\times(\bm{m}_3-\bm{m}_1)\big)\cdot(\bm{m}_4 - \bm{m}_1)\big)\big|.
\ee
This is nothing but the volume of the tetrahedron.

\begin{figure}[ht]
\begin{center}
\includegraphics[width=11cm]{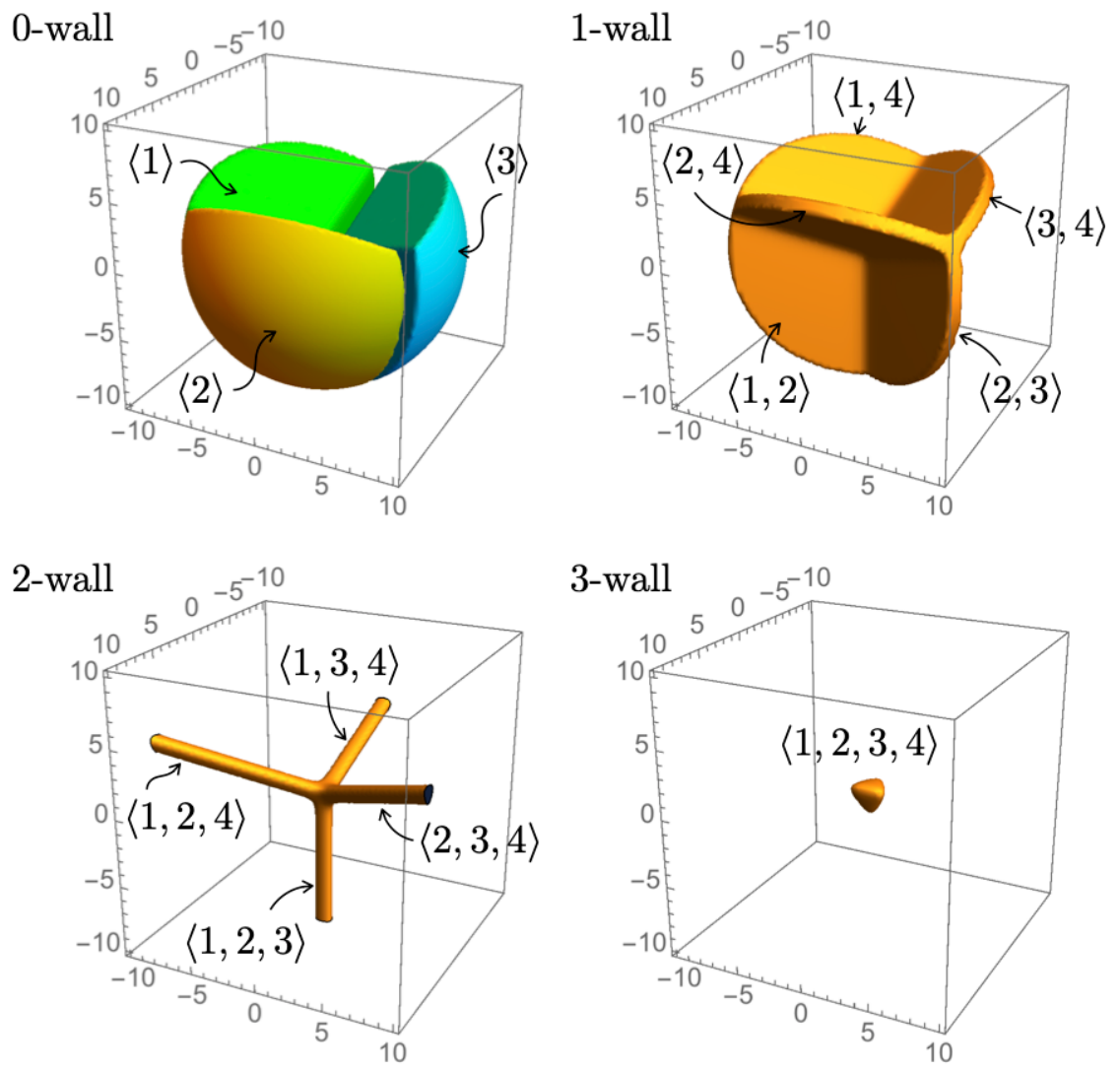}
\caption{The real space $\mathbb{R}^3$: isosurfaces of the topological charge 
densities ${\cal W}_d$ for $d$-walls ($d=0,1,2,3$) are shown. We only depict the densities
within the sphere of the radius $10$.}
\label{fig:Nf4}
\end{center}
\end{figure}

In Fig.~\ref{fig:Nf4},  
we show the topological charge densities ${\cal W}_d$ 
for the regular tetrahedron given in Fig.~\ref{fig:Nf_4_grid}.
The top-left panel shows the vacuum (0-wall) domains 
($\left<1\right>$ = green, 
$\left<2\right>$ = yellow, 
$\left<3\right>$ = cyan,
$\left<1\right>$ = hidden). 
At boundaries between two vacuum (0-wall) domains, there are domain walls (1-walls), 
as shown in the top-right panel. The plot shows an isosurface of the sum of three
1-wall charge densities ${\cal W}_1^{(x)}+{\cal W}_1^{(y)}+{\cal W}_1^{(z)}$.
Similarly, the bottom-left panel shows an isosurface of the sum of the three 2-wall charge
densities ${\cal W}_2^{(xy)}+{\cal W}_2^{(yz)}+{\cal W}_2^{(zx)}$.
Finally, the bottom-right panel shows an isosurface of the 3-wall charge density
${\cal W}_3^{(xyz)}$. The figures clearly show that the generalized topological charge
defined in Eq.~(\ref{eq:W_d}) is appropriate to describe the codimension $d$ structure
in the solution.

\begin{figure}[ht]
\begin{center}
\includegraphics[width=15cm]{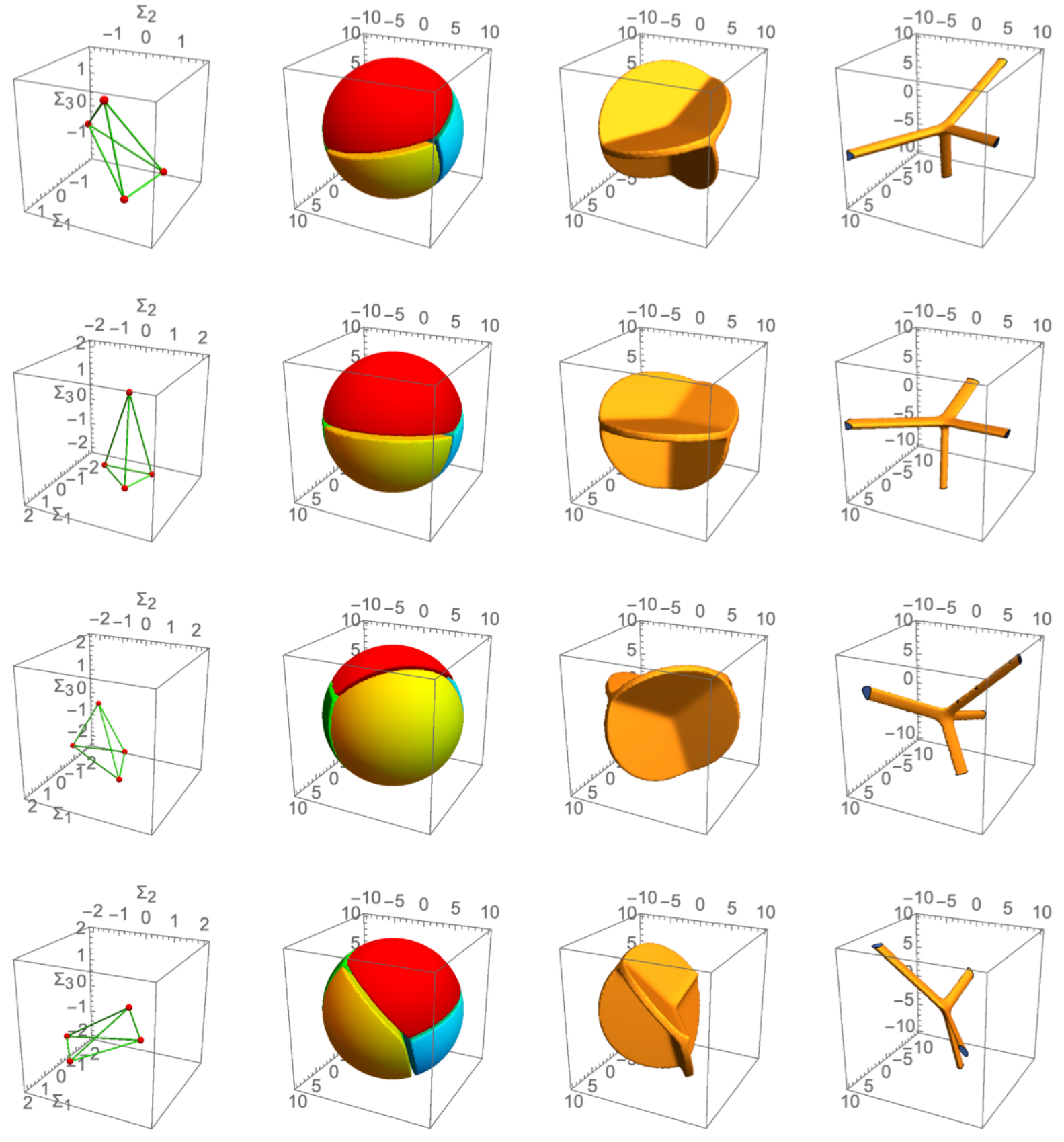}
\caption{Four examples of the BPS solutions: The left-most panel shows the grid diagrams,
and the second, third, and fourth from left show the corresponding ${\cal W}_d$ 
$d=0,1,2$. We only depict the densities
within the sphere of the radius $10$.}
\label{fig:Nf4_gene}
\end{center}
\end{figure}

To close this subsection, 
we again emphasize that we have chosen the symmetric arrangement of the masses corresponding 
to a regular tetrahedron in Fig.~\ref{fig:Nf_4_grid} just for simplicity, 
but the exact solution has been obtained for arbitrary mass arrangement.
In order to make this point clearer, we show in Fig.~\ref{fig:Nf4_gene} 
four solutions for randomly chosen
tetrahedrons.

%%%%%%%%%%%%%
\subsection{$N =5$: Dipyramid: Minimal domain wall networks}

\begin{figure}[ht]
\begin{center}
\includegraphics[width=14cm]{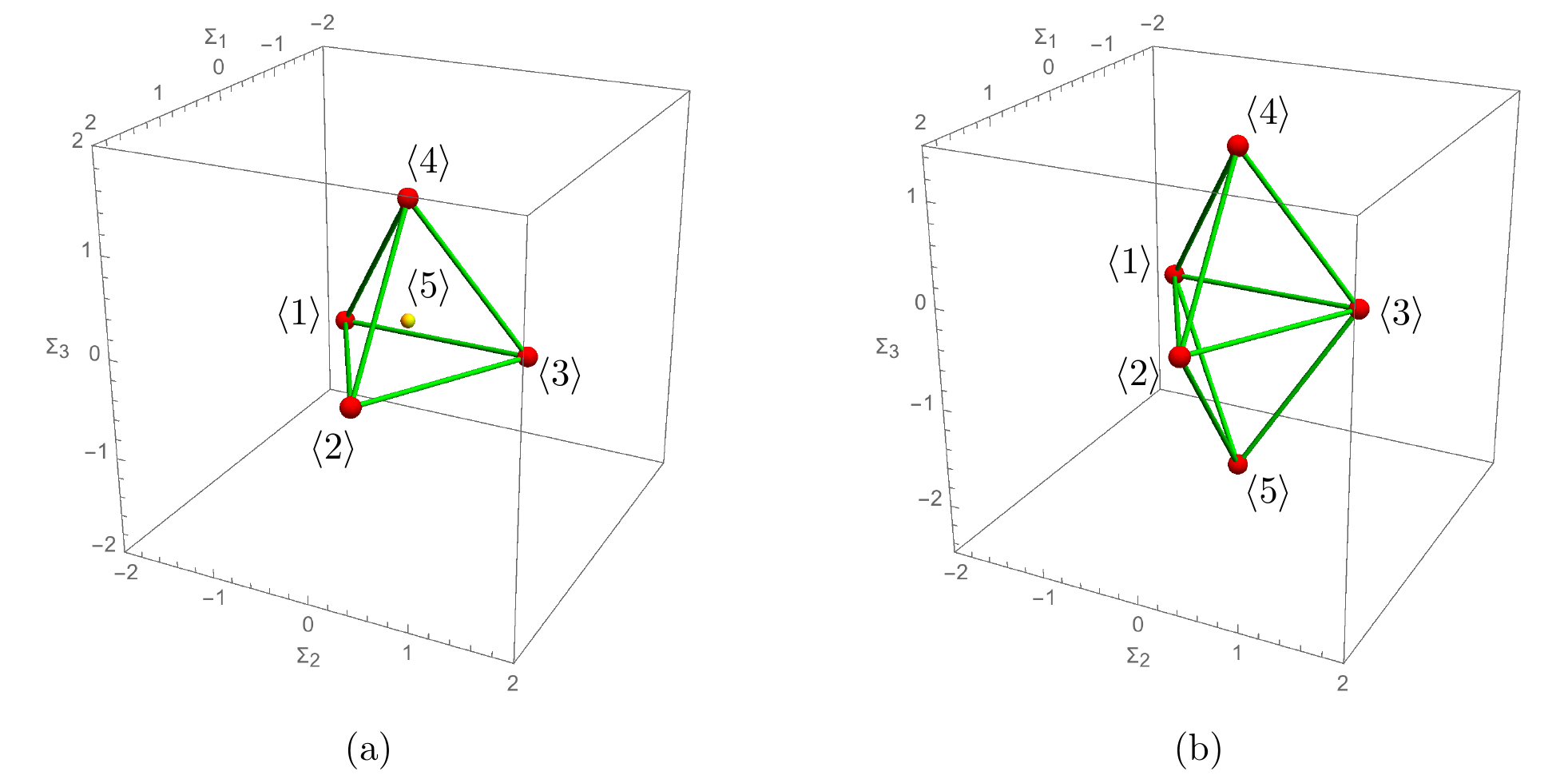}
\caption{The grid diagrams ($\bm{\Sigma}$ space): 
(a) and (b) show the two different
patterns of the five vacua in $N =5$.}
\label{fig:Nf_5_grid}
\end{center}
\end{figure}

Let us next consider a model 
admitting a non-planar network structure of domain walls. 
The minimal number for this is $N =5$ for which 
there is one additional vacuum $\left<5\right>$ compared to the model with $N =4$.
According to where we put the fifth vacuum, the resulting network structures are
classified into two types as shown in Fig.~\ref{fig:Nf_5_grid} (a) and (b).
The feature of (a) is the presence of an inner vacuum $\left<5\right>$ in the
tetrahedron $\left<1,2,3,4\right>$. 
On the other hand, Fig.~\ref{fig:Nf_5_grid}(b) is a dipyramid for which 
the fifth vacuum $\left<5\right>$ is
placed outside the tetrahedron $\left<1,2,3,4\right>$. It is a convex polytope and has no
inner vacua.

Let us start with the former case 
of the tetrahedron with the inner vacuum. 
We take the same four vacua
defined in Eqs.~(\ref{eq:M_tetra_1})--(\ref{eq:M_tetra_4}) for simplicity, 
and the fifth one is set as
\be
\bm{m}_5 = (0,0,0).
\ee
The most generic moduli matrix up to the three translations in $\mathbb{R}^3$ and
the $V$-transformation (\ref{eq:V-transf}) is given by
\be
H_0 = (1,1,1,1,e^{a_5}).
\label{eq:MM_Nf5a}
\ee
Then $\Omega$ depends on the one moduli parameter $a_5$ as
\be
\Omega = \sum_{A=1}^4 e^{2\tilde{\bm{m}}_A \cdot \bm{x}} + e^{2a_5}e^{2\tilde{\bm{m}}_5 \cdot \bm{x}}.
\ee
\begin{figure}[ht]
\begin{center}
\includegraphics[width=15cm]{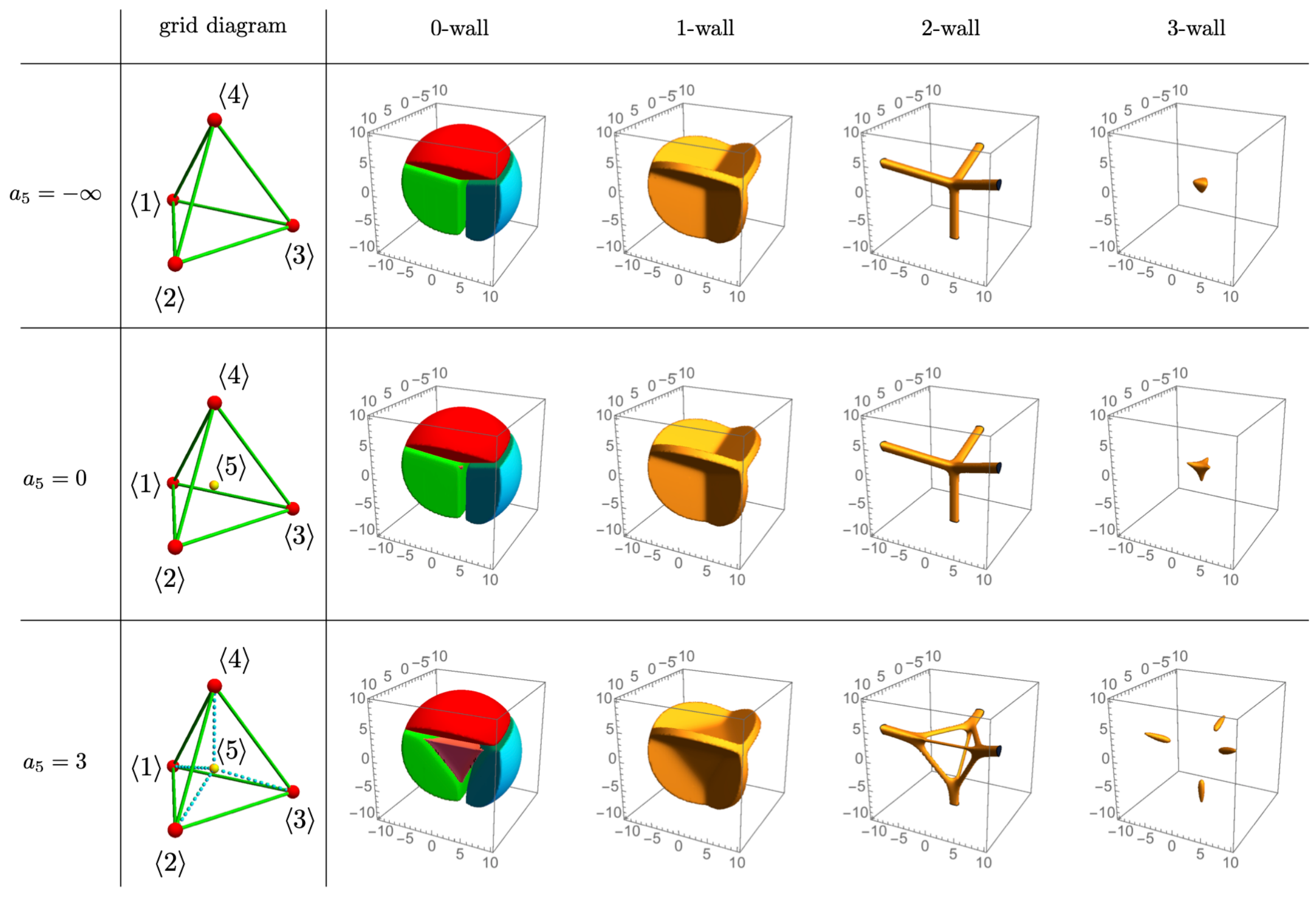}
\caption{Exact solutions in three typical branches 
for the grid diagram in Fig.~\ref{fig:Nf_5_grid}(a).
Isosurfaces of the topological charge 
densities ${\cal W}_d$ for $d$-walls ($d=0,1,2,3$) are shown.}
\label{fig:Nf5a}
\end{center}
\end{figure}
From this expression, we understand that the factor $e^{a_5}$ controls the strength
(the weight)
of the fifth vacuum $\left<5\right>$  relative to the rests. 
When $a_5 \to -\infty$, the $\left<5\right>$ domain disappears.
The $\left<5\right>$ domain extends as $e^{a_5}$ increases. This moduli dependence of the
configuration can be seen 
in Fig.~\ref{fig:Nf5a} in which we have shown three examples with $a_5 = -\infty$, $0$, and $3$.
The large distance behaviors are the same for all these cases. However,
a significant difference emerges around the origin. Reflecting the inner vacuum $\left<5\right>$
inside the tetrahedron, a compact vacuum domain of $\left<5\right>$ emerges by increasing $a_5$.
The whole structure of the solution is the following. There  is one inner vacuum bubble 
$\left<5\right>$ surrounded by the semi-infinite four vacuum domains $\left<1\right>$,
$\left<2\right>$, $\left<3\right>$, and $\left<4\right>$.
There are six semi-infinite 1-walls corresponding to the six outer edges 
 ($\left<1,2\right>$, $\left<1,3\right>$, $\left<1,4\right>$, $\left<2,3\right>$,
$\left<2,4\right>$, and $\left<3,4\right>$), 
and four compact 1-walls corresponding to the four inner edges
($\left<1,5\right>$, $\left<2,5\right>$, $\left<3,5\right>$, $\left<4,5\right>$), see 
the third row of Fig.~\ref{fig:Nf5a}.
There are four semi-infinite 2-walls and the six compact 2-walls corresponding to
the four faces ($\left<1,2,3\right>$, $\left<1,2,4\right>$, $\left<1,3,4\right>$,
$\left<2,3,4\right>$) and the six inner triangles ($\left<1,2,5\right>$, $\left<2,3,5\right>$,
$\left<1,3,5\right>$, $\left<1,4,5\right>$, $\left<2,4,5\right>$, $\left<3,4,5\right>$), respectively.
Finally, there are four 3-walls which correspond to the four sub-tetrahedrons
($\left<1,2,4,5\right>$, $\left<2,3,4,5\right>$, $\left<1,3,4,5\right>$, $\left<1,2,3,5\right>$).
The network structure in the real $\mathbb{R}^3$ space
can be best seen in the plot of the 2-walls as shown in Fig.~\ref{fig:Nf5a}.

\begin{figure}[t]
\begin{center}
\includegraphics[width=15cm]{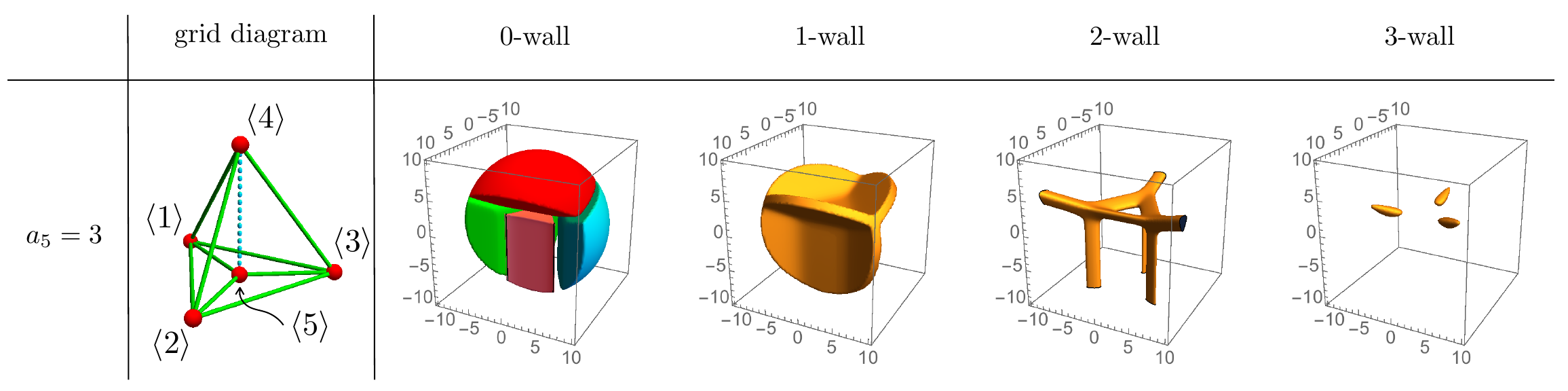}
\caption{Exact solution
for the fine-tuned grid diagram.
Isosurfaces of the topological charge 
densities ${\cal W}_d$ for $d$-walls ($d=0,1,2,3$) are shown.}
\label{fig:Nf5d}
\end{center}
\end{figure}

Next, let us place the fifth vacuum $\left<5\right>$ on the plane including the
bottom triangle $\left<1,2,3\right>$, say
\be
\bm{m}_5 = \left(0,0,-\frac{1}{\sqrt 6}\right).
\ee
The moduli matrix is the same as the one in Eq.~(\ref{eq:MM_Nf5a}).
When $e^{a_5}$ is sufficiently large, the configuration becomes a mixture of 
planar and non-planar structures, as shown in Fig.~\ref{fig:Nf5d} for $a_5 = 3$.
The planar structure comes from the bottom triangle $\left<1,2,3\right>$ which are
divided into three sub-triangles $\left<1,2,5\right>$, $\left<2,3,5\right>$, and
$\left<1,3,5\right>$. As a consequence, there are six parallel semi-infinite 1-walls
corresponding to $\left<1,2\right>$, $\left<1,3\right>$, $\left<1,5\right>$,
$\left<2,3\right>$, $\left<2,5\right>$, and $\left<3,5\right>$. On the other hand,
the non-planar structure is originated from the vacuum $\left<4\right>$ which is placed
off the bottom triangle.
In addition to the semi-inifnite 1-walls corresponding to
the edges $\left<1,4\right>$, $\left<2,4\right>$, $\left<3,4\right>$, there is the
compact 1-wall of $\left<4,5\right>$. Comparing the 2-wall and 3-wall structures 
in Figs.~\ref{fig:Nf5a} and \ref{fig:Nf5d}, we see that 
the lower 3-wall is pushed down towards infinity, which corresponds to the fact
that  the sub-tetrahedron $\left<1,2,3,5\right>$ gets squashed flat.

\begin{figure}[t]
\begin{center}
\includegraphics[width=15cm]{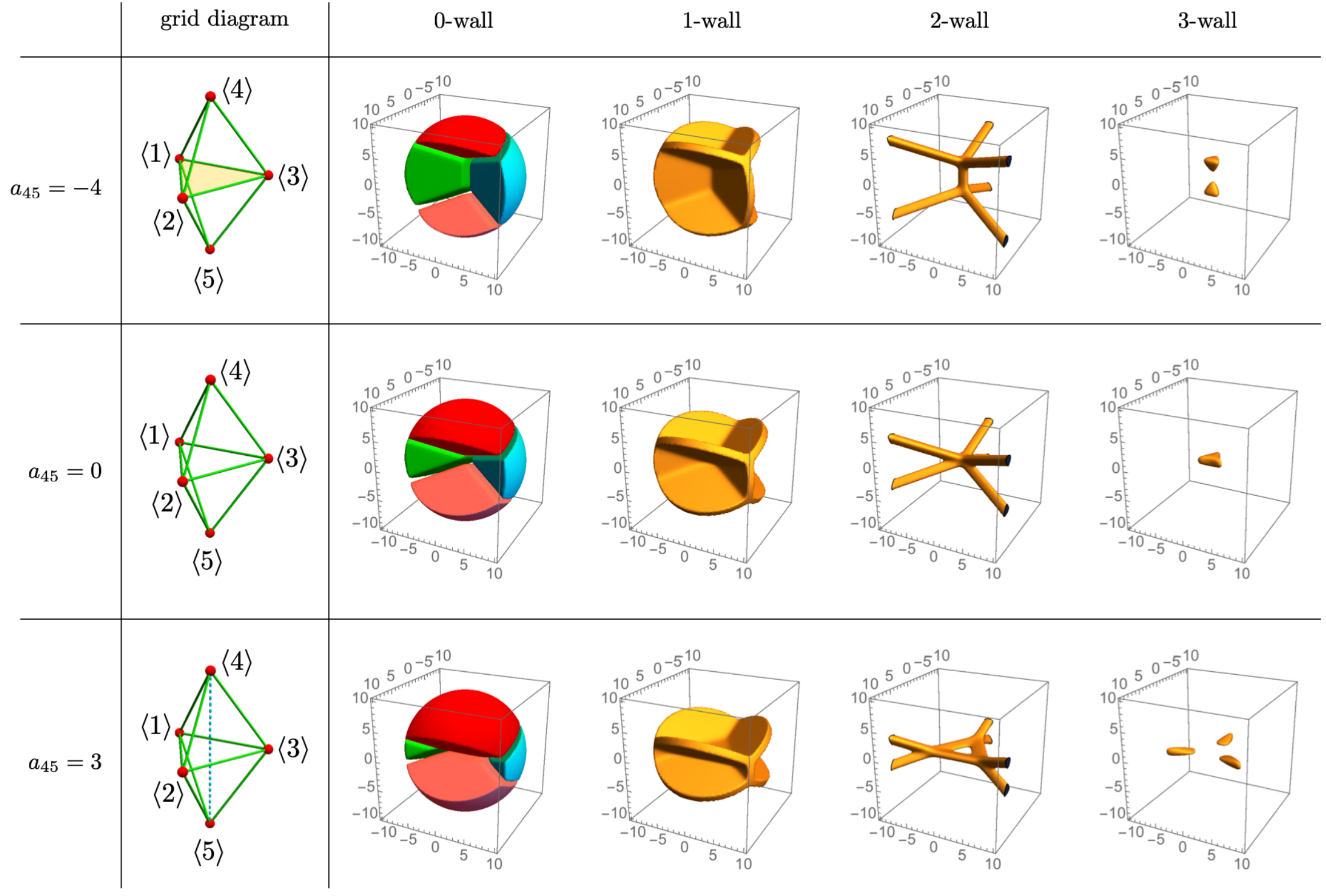}
\caption{Exact solutions in three typical branches 
for the grid diagram in Fig.~\ref{fig:Nf_5_grid}(b).
Isosurfaces of the topological charge 
densities ${\cal W}_d$ for $d$-walls ($d=0,1,2,3$) are shown.}
\label{fig:Nf5b}
\end{center}
\end{figure}

Let us next study the second type with the dypyramid structure as Fig.~\ref{fig:Nf_5_grid}(b),
which can be obtained by  further pushing down the vacuum $\left<5\right>$ from
the last case in Fig.~\ref{fig:Nf5d}.
As before, the first four masses $\bm{m}_1$, $\bm{m}_2$, 
$\bm{m}_3$, and $\bm{m}_4$ are intact. Then the fifth vacuum is placed at the opposite to
$\left<4\right>$ with respect to the triangle $\left<1,2,3\right>$ as
\be
\bm{m}_5 = \left(0,0,-\frac{5}{\sqrt{6}}\right).
\ee
A useful choice of the moduli matrix with one moduli parameter $a_{45}$ is given by
\be
H_0 = (1,1,1,e^{a_{45}},e^{a_{45}}).
\ee
The parameter $a_{45}$ controls the weight of the vacua $\left<4\right>$
and $\left<5\right>$ relative to that of $\left<1\right>$, $\left<2\right>$, and $\left<3\right>$.
We show three typical solutions with $a_{45} = -4,0, 3$ in Fig.~\ref{fig:Nf5b}.

The first row corresponds to $a_{45} = -4$. The $\left<4\right>$ and $\left<5\right>$
domains are relatively weak whereas the domains of $\left<1\right>$, $\left<2\right>$,
and $\left<3\right>$ stick out and directly meet to form a 2-wall corresponding to
the inner triangle $\left<1,2,3\right>$. The dypyramid is divided into the upper
and the lower (upside down) tetrahedrons. Correspondingly, the configurations (1-, 2-, 3-walls)
are constructed by joining two tetrahedral solutions at the $z=0$ plane.

The middle row of Fig.~\ref{fig:Nf5b} corresponds to $a_{45}=0$ where the strengths of all the vacua
comparable. Namely, all the vacuum domains meet at the origin.
Accordingly, the inner 2-wall corresponding to the triangle $\left<1,2,3\right>$ disappears.

As increasing $a_{45}$ further, the vacuum domains $\left<4\right>$ and $\left<5\right>$
expand, and directly meet to create new 1-wall $\left<4,5\right>$, see
the third row of Fig.~\ref{fig:Nf5b}. In this case, the whole dypyramid can be thought
of as the sum of three sub-tetrahedrons $\left<1,2,4,5\right>$, $\left<1,3,4,5\right>$,
and $\left<2,3,4,5\right>$. This can be seen in the 2- and 3-walls depicted
in the third row of Fig.~\ref{fig:Nf5b}.

\medskip
\begin{figure}[t]
\begin{center}
\includegraphics[width=15cm]{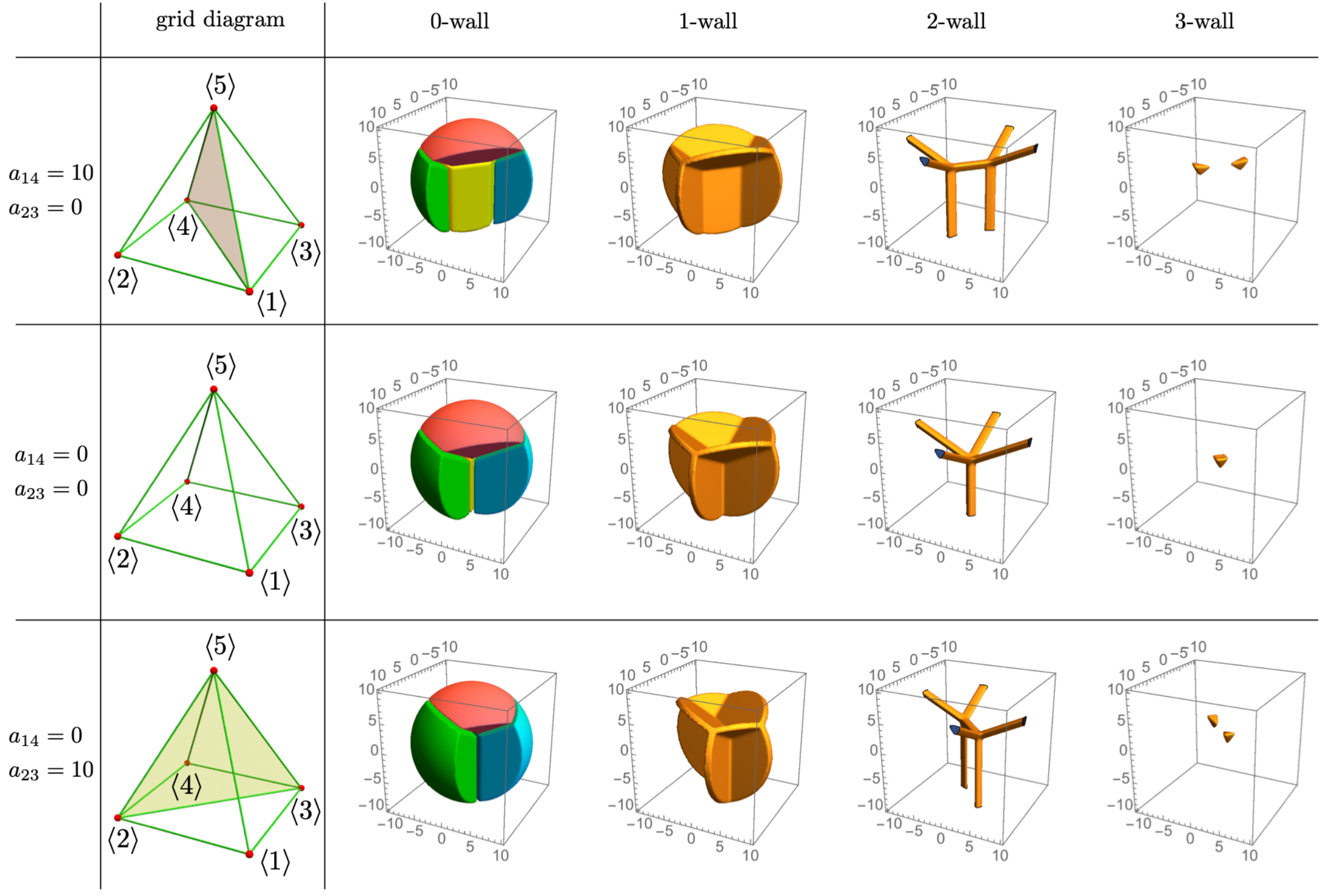}
\caption{Exact solutions in three typical branches 
for the fine-tuned grid diagram.
Isosurfaces of the topological charge 
densities ${\cal W}_d$ for $d$-walls ($d=0,1,2,3$) are shown.}
\label{fig:Nf5c}
\end{center}
\end{figure}
Finally, we consider a rare case that the four vacua are placed on a plane similarly
to Fig.~\ref{fig:Nf5d}. We now put the four points on a plane so that they form
a square, see Fig.~\ref{fig:Nf5c}. Namely, the five vacua form a square pyramid. 
As an example, our choice is
\be
\bm{m_1} &=& \left(2,2,0\right),\\
\bm{m_2} &=& \left(2,-2,0\right),\\
\bm{m_3} &=& \left(-2,2,0\right),\\
\bm{m_4} &=& \left(-2,-2,0\right),\\
\bm{m_5} &=& \left(0,0,4\right).
\ee 
The moduli matrix useful for this case is
\be
H_0 = (e^{a_{14}},e^{a_{23}},e^{a_{23}},e^{a_{14}},1),
\ee
where $a_{14}$ corresponds to the weights of $\left<1\right>$ and $\left<4\right>$
whereas $a_{23}$ corresponds to the weights of $\left<2\right>$ and $\left<3\right>$.\footnote{The 
parameters $a_{14}$ and $a_{23}$ are not physically independent moduli
due to the $V$-transformation (\ref{eq:V-transf}).}
We show three typical solutions with $(a_{14},a_{23}) = (10,0),\ (0,0),\ (0,10)$ 
in Fig.~\ref{fig:Nf5c}.

In  the first row of Fig.~\ref{fig:Nf5c},
the first solution with $(a_{14},a_{23}) = (10,0)$ is shown.
Since the weights of $\left<1\right>$ and $\left<4\right>$ are greater than those
of $\left<2\right>$ and $\left<3\right>$, the former domains stick out and directly meet to
form the 1-wall $\left<1,4\right>$.

When $(a_{14},a_{23}) =  (0,0)$, all the vacua have the same influences, so that they meet
at a point, see the middle row of Fig.~\ref{fig:Nf5c}. 

When $(a_{14},a_{23}) = (0,10)$, 
the influence of $\left<2\right>$ and $\left<3\right>$ are the maximum, so that their
domains contact and form the 1-wall $\left<2,3\right>$ as shown in the third row of Fig.~\ref{fig:Nf5c}.

A difference of the two cases $(a_{14},a_{23}) = (10,0)$ and $(0,10)$ is just difference
of dividing the square pyramid into two tetrahedrons.
If we only look at the square face, it is a transition between the s- and t-channels found
for a planar 1-wall network, discussed in Ref.~\cite{Eto:2005cp}. 
Therefore, this sort of cutting the square pyramid into two tetrahedrons is 
a three dimensional analog of the transition between the s- and t-channels in two dimensions. One can be convinced by looking at at the 2-wall (but not 1-wall)
configurations in Fig.~\ref{fig:Nf5c} from above (or bottom).

%%%%%%%%%%%%
\subsection{$N =6$: Octahedron}

Let us further increase the number of vacua, namely $N  = 6$.
There are six vacua. There are three different cases according to
the convex polyhedra made by connecting the vacua. The first 
case is a tetrahedron with two inner vertices, the second
is a dipyramid with a inner vertex, and the third is a octahedron without
inner vertices. 

\begin{figure}[t]
\begin{center}
\includegraphics[width=13cm]{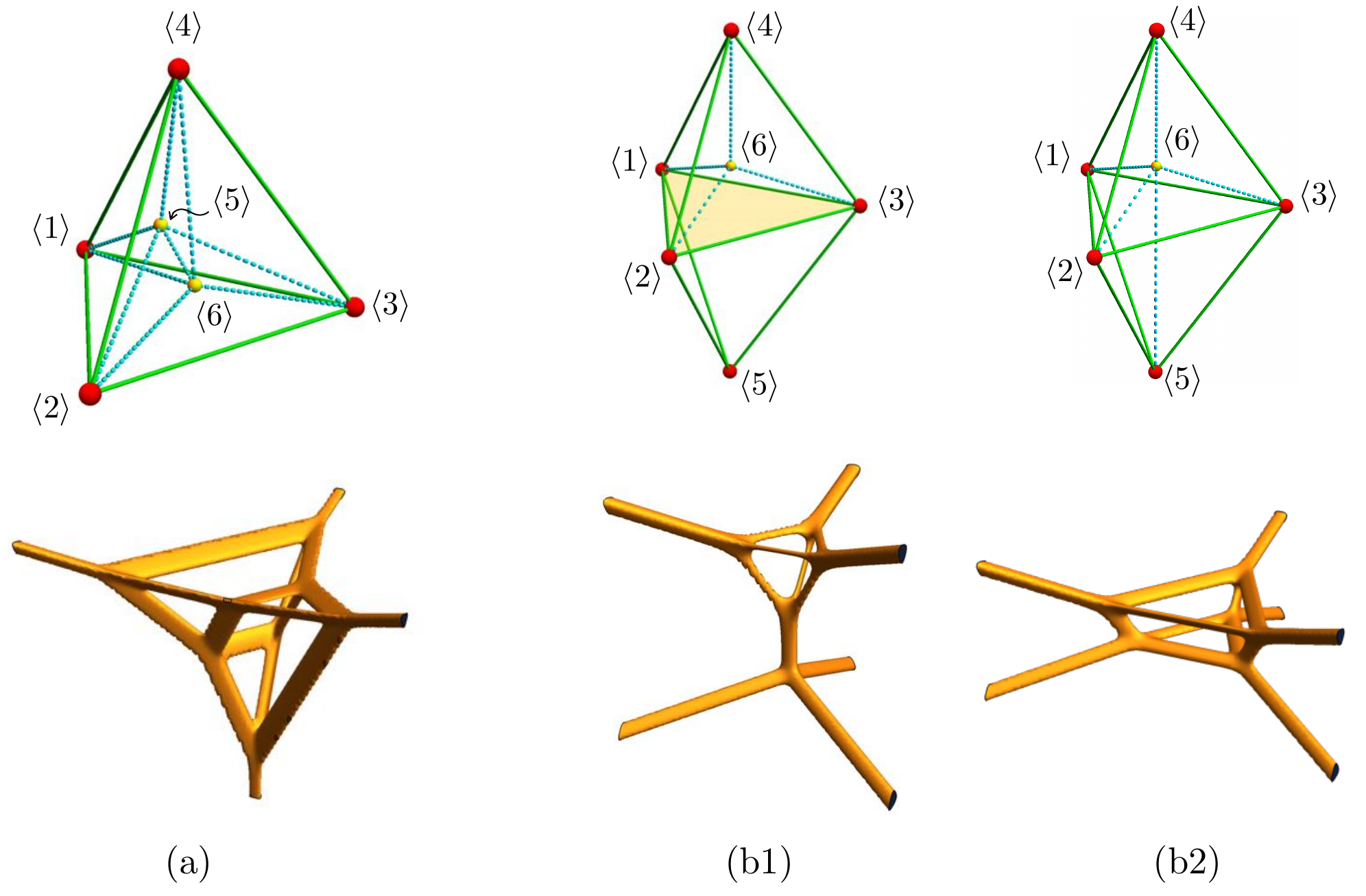}
\caption{Exact solutions of domain wall networks for the model with $N =6$.
The first line shows the grid diagrams and the second line shows isosurfaces of
${\cal W}_2$ for the corresponding solutions.
(a) The grid diagram is a tetrahedron and it has two inner vertices.
(b1) and (b2) are two different type of the networks for the dipyramid-form grid
diagram with one inner vertex. (a) has two vacuum bubbles, and (b1) and (b2) have
one vacuum bubble.}
\label{fig:Nf6b}
\end{center}
\end{figure}

The tetrahedral case is shown in Fig.~\ref{fig:Nf6b}(a).
The whole tetrahedron is divided into eight sub-tetrahedrons 
$\left<1,2,3,6\right>$,
$\left<1,2,4,5\right>$,
$\left<1,3,4,5\right>$,
$\left<2,3,4,6\right>$,
$\left<1,2,5,6\right>$,
$\left<2,3,5,6\right>$,
$\left<3,4,5,6\right>$,
$\left<1,4,5,6\right>$.
Correspondingly, there exist eight 3-walls (junctions of four 2-walls).
The two inner vertices give rise to the two vacuum bubbles
as shown in the bottom figure of Fig.~\ref{fig:Nf6b}(a).
The concrete moduli matrix for Fig.~\ref{fig:Nf6b}(a) is
\be
H_0 = \left(1,1,1,1, e^{a_5}, e^{a_6}\right),
\ee
with $a_5 = a_6 = 10$. The moduli parameter $a_5$ ($a_6$) controls size of the
bubble of $\left<5\right>$ ($\left<6\right>$).

For the dipyramid type case, there are two branches according to how to divide it
into sub-tetrahedrons. The first branch is shown in Fig.~\ref{fig:Nf6b}(b1) in which 
the whole dipyramid is divided into five tetrahedrons,
$\left<1,2,3,6\right>$, 
$\left<1,2,4,6\right>$,
$\left<1,3,4,6\right>$,
$\left<2,3,4,6\right>$,
and
$\left<1,2,3,5\right>$. It can be also regarded as the octahedron made of two tetrahedrons
bonded at the surface $\left<1,2,3\right>$. The upper tetrahedron has the inner vertex whereas
the bottom one has no inner vertices. The resulting 2-wall wireframe shown
in the lower figure of Fig.~\ref{fig:Nf6b}(b1) can be indeed obtained by
connecting the 2-walls of Fig.~\ref{fig:Nf4} and the third row of Fig.~\ref{fig:Nf5a}.

On the other hand, the whole dipyramid is divided into six subtetrahedrons
$\left<1,2,4,6\right>$,
$\left<2,3,4,6\right>$,
$\left<1,3,4,6\right>$,
$\left<1,2,5,6\right>$,
$\left<2,3,5,6\right>$,
and $\left<1,3,5,6\right>$
in the second branch, see Fig.~\ref{fig:Nf6b}(b2).

The useful moduli matrix for describing this transition turns out to be
\be
H_0 = \left(e^{a_{123}},e^{a_{123}},e^{a_{123}},1,1,e^{a_6}\right).
\ee
The moduli parameter $a_{123}$ controls strength of the vacua $\left<1\right>$,
$\left<2\right>$, and $\left<3\right>$. In other words, it is related to distance between the 3-walls
$\left<1,2,3,5\right>$ and $\left<1,2,3,6\right>$. The other $a_6$
controls the size of the bubble $\left<6\right>$. We choose $(a_{123},a_6) = (10,10)$
in Fig.~\ref{fig:Nf6b}(b1), and $(a_{123},a_6) = (-2,4)$ in Fig.~\ref{fig:Nf6b}(b2).

\begin{figure}[t]
\begin{center}
\includegraphics[width=14cm]{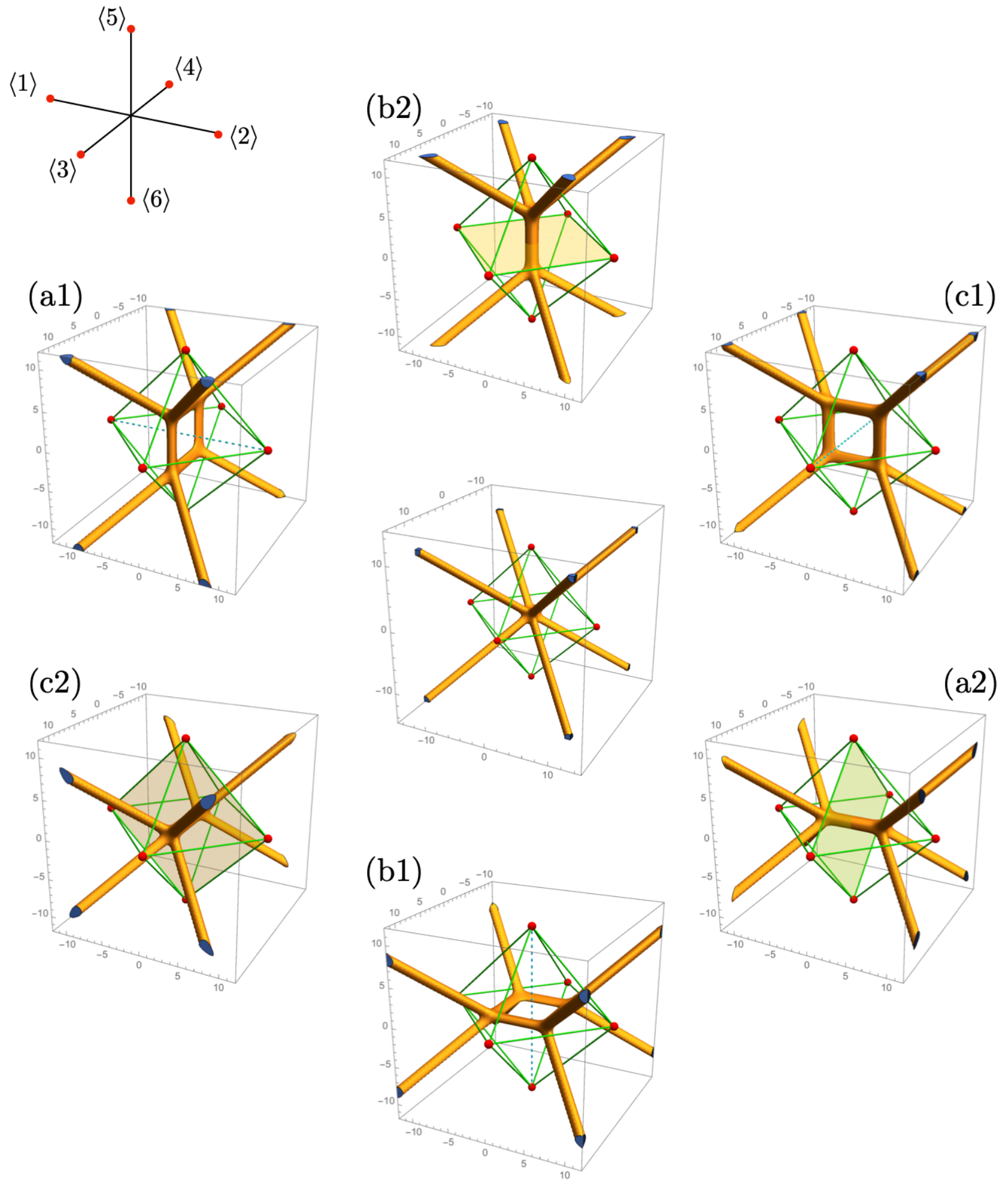}
\caption{Exact solutions of domain wall network in the octahedral grid diagram ($N =6$)
without inner vertices. We superpose the two kinds of graphs, the one is the ${\cal W}_2$
in the $\bm{x}$-space and the other is the grid diagram in the $\bm{\Sigma}$-space.
There are three orthogonal directions to deform the network. The figure at the center shows
the network where all the six vacua have the same influences. (a1) [(a2)]
is obtained when the weight of $\left<1\right>$ and $\left<2\right>$ is larger (smaller)
than the other vacua. Similarly, (b1) and (b2) [(c1) and (c2)] transit by controlling the relative
influence of $\left<5\right>$ and $\left<6\right>$ [$\left<3\right>$ and $\left<4\right>$].}
\label{fig:Nf6a}
\end{center}
\end{figure}
Finally, we show the octahedral case in Fig.~\ref{fig:Nf6a}. For simplicity, we set the
eight masses $\bm{m}_A$ at vertices of a regular octahedron.
This configuration can be understood as follows. 
To understand this configuration, we arrange the six vacua to the following three pairs:
$\left<1\right>$-$\left<2\right>$, $\left<3\right>$-$\left<4\right>$, and
$\left<5\right>$-$\left<6\right>$.
Then, the appropriate moduli matrix is
\be
H_0 = \left(e^{a_{12}}, e^{a_{12}}, e^{a_{34}}, e^{a_{34}}, e^{a_{56}}, e^{a_{56}} \right),
\ee
where only two among $a_{12}, a_{34}, a_{56}$ are independent.
When the weight of the vacua $\left<1\right>$ and $\left<2\right>$ are larger
than the other vacua, the vacua $\left<1\right>$ and $\left<2\right>$ directly meet
to form the 1-wall $\left<1,2\right>$. It corresponds to (a1) of Fig.~\ref{fig:Nf6a} in which
the vertices $\left<1\right>$ and $\left<2\right>$ are connected by the dashed segment,
and at the same time there is an inner loop of the 2-wall penetrated by the segment.
As the weight relatively decreases, the loop shrinks. When $a_{12} = a_{34} = a_{56}$,
all the vacua are equivalent and there are no 2-wall loops, as shown in the center panel
of Fig.~\ref{fig:Nf6a}. If we further reduce $a_{12}$ as
$a_{12} < a_{34} = a_{56}$, the four vacua $\left<3\right>$, $\left<4\right>$, $\left<5\right>$,
and $\left<6\right>$ stick out and directly meet each other, see Fig.~\ref{fig:Nf6a}(a2).
The same can be said to the other pairs $\left<3\right>$-$\left<4\right>$ and
$\left<5\right>$-$\left<6\right>$. When $a_{56} > a_{12}=a_{34}$ ($a_{56} < a_{12}=a_{34}$), 
we have (b1) [(b2)] of Fig.~\ref{fig:Nf6a}.
When $a_{34} > a_{12}=a_{56}$ ($a_{34} < a_{12}=a_{56}$), 
we have (c1) [(c2)] of Fig.~\ref{fig:Nf6a}. We can rephrase these as follows. The configurations
of (a1), (b1), and (c1) correspond to the decomposition of the octahedron into four sub-tetrahedrons.
On the other hand, (a2), (b2), and (c2) correspond to the decomposition of the octahedron into two square pyramids.

%%%%%%%%%%%%%%%%% S E C T I O N 6 %%%%%%%%%%%%%%%%%%

\section{Platonic, Archimedean, Catalan, and Kepler-Poinsot vacuum bubbles}
\label{sec:platon}

In the previous sections, we have seen several examples in which the configurations
have vacuum bubbles, see Figs.~\ref{fig:Nf5a} and \ref{fig:Nf6b}. They can appear
when the grid diagrams which are generally convex polytopes have inner vertices.
The number of the inner vertices is equal to the number of the vacuum bubbles.
The size of a vacuum bubble is controlled by a modulus. The bubble is sometimes invisibly small, 
and we need to take the corresponding moduli parameter sufficiently large to
broaden the bubble.

\subsection{The Platonic vacuum bubbles}
Here, we are interested in the shape of the vacuum bubble.
In Fig.~\ref{fig:Nf5a} we met the bubble which is a regular tetrahedron.
There are two conditions for a regular tetrahedral bubble to exist. 
One is that the grid diagram is a regular tetrahedron, and the other
is that the inner vertex is located at the center of the regular tetrahedron.
If we relax either or both conditions, the bubble deforms accordingly.

\begin{figure}[t]
\begin{center}
\includegraphics[width=12cm]{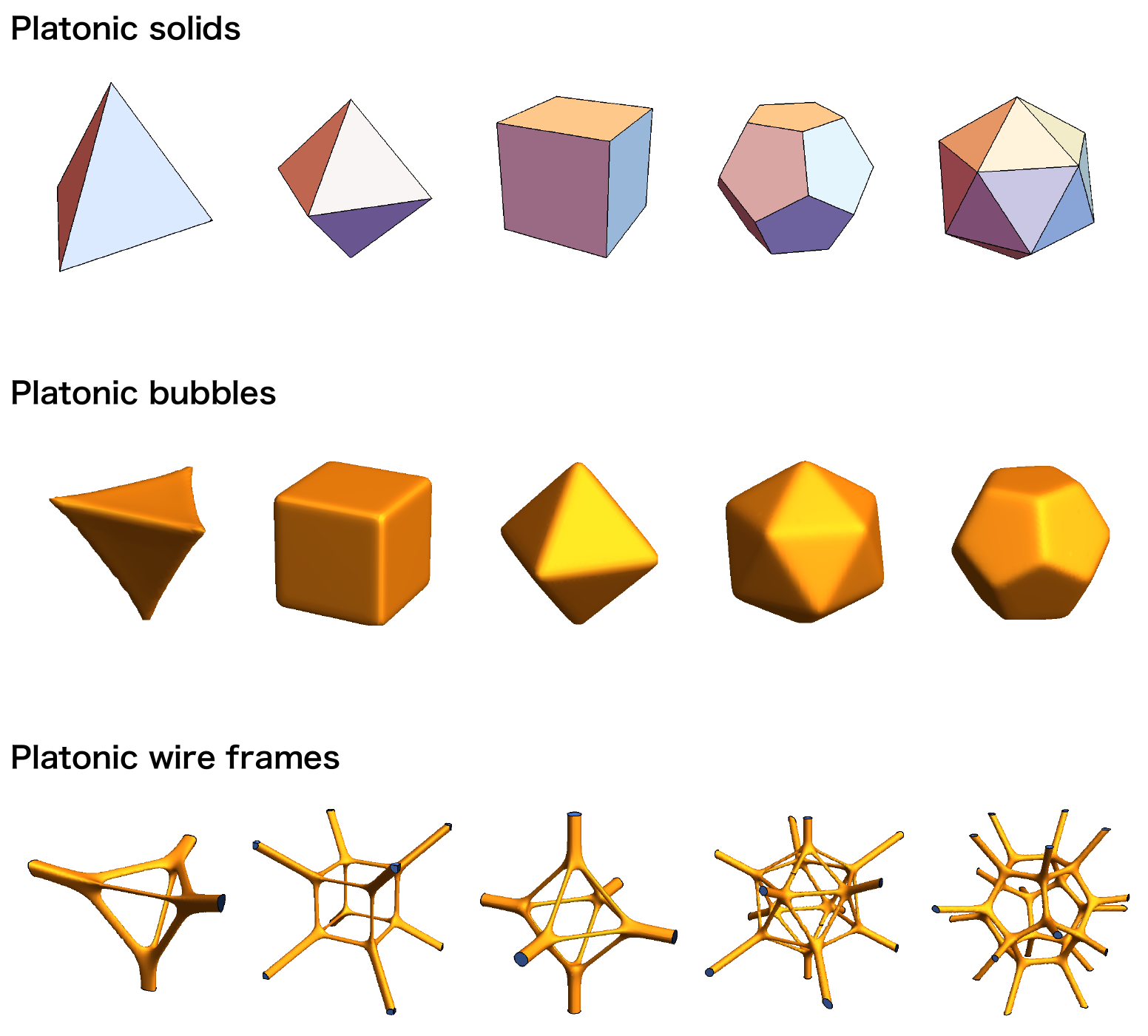}
\caption{The top row shows five grid diagrams which are congruent with the Platonic solids.
The inner vertices at the center are hidden. The second row shows the vacuum bubbles for 
each grid diagrams, and the third row shows the 2-wall charge densities ${\cal W}_2$. The
bubble shape and the grid diagram are dual each other.}
\label{fig:Platonic_bubble}
\end{center}
\end{figure}

Now, one notices that the two regular tetrahedrons in the grid diagrams and
in the real space in Fig.~\ref{fig:Nf5a} are upside-down. 
This should be so, since the each face of the bubble is perpendicular to the inner edge
of the grid diagram. This can be rephrased as follows. The shape of the vacuum bubble
is equivalent to a shape obtained by exchanging the outer vertices and faces 
of the grid diagram. Such a polyhedron is called {\it dual} of the original polyhedron.
Since the regular tetrahedron is known to be self-dual, the grid diagram and the vacuum bubble
are both  regular tetrahedrons. We show the regular tetrahedron as the grid diagram,
the dual tetrahedron as the vacuum bubble, and an isosurface of the 2-wall density
in the left-most column in Fig.~\ref{fig:Platonic_bubble}. 

To understand the duality better,
let us next consider domain wall network in a shape of  
the other convex regular polyhedra: octahedron, cube, icosahedron, and dodecahedron.
Namely, we consider the grid diagrams which are congruent with the Platonic solids, and put
an inner vertex at their centers. 

For example, we consider $N =6+1$ for the octahedron with
the inner vertex. Let the first six flavors ($A=1,2,\cdots,6$) 
correspond to the vertices of the octahedron,
and the seventh flavor ($A=7$) to the inner vertex. 
To get a vacuum bubble of a desired shape,
we consider the moduli matrix $H_0 = (1,1,1,1,1,1, e^{a_7})$. Then we take a sufficiently 
large value for $e^{a_7}$ to broaden the vacuum bubble $\left<7\right>$. The result is shown
in the second column from the left of Fig.~\ref{fig:Platonic_bubble}.
We indeed observe that the resultant vacuum bubble takes a shape of a regular cube,  
dual to a regular octahedron. 

Similarly, we take $N =8+1$ for a regular cube plus an inner vertex,
$N  = 20+1$ for a regular dodecahedron plus an inner vertex, and $N =12+1$ for a regular
icosahedron plus an inner vertex.
We find that the shapes of the vacuum bubbles are a regular octahedron, icosahedron, and
dodecahedron for the regular cube, dodecahedron, and icosahedron, respectively,
see Fig.~\ref{fig:Platonic_bubble}.
Thus, putting an inner vertex at the centers 
of the Platonic solids and broadening the corresponding
vacuum give us the regular bubble polyhedra 
dual to the Platonic solids as the grid diagrams.

Here, we equally set all the weights of the outer vertices to be 1 to have the regular
polyhedrons. Of course, if we choose a generic moduli matrix, the vacuum bubbles deform accordingly,
as mentioned at the beginning of this subsection.

\subsection{The Archimedean and Catalan vacuum bubbles}
Let us also test the idea to the other classic polyhedra, the thirteen Archimedean polyhedra
and their duals called the Catalan polyhedra.
The Archimedean polyhedra are uniform polyhedra which are vertex-transitive.
Therefore, the Catalan polyhedra are face-transitive.

\begin{figure}[th]
\begin{center}
\includegraphics[width=15cm]{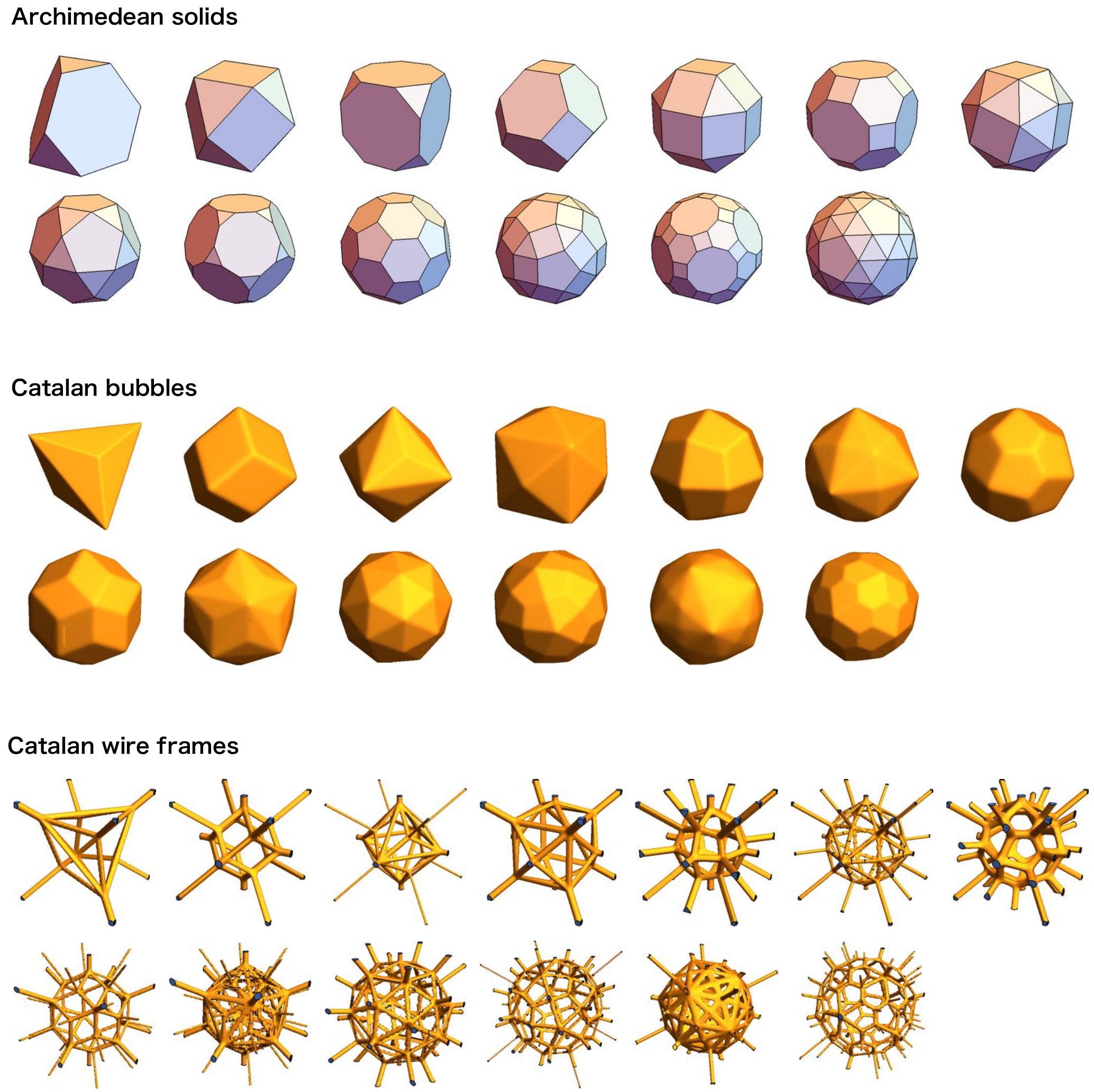}
\caption{The top row shows thirteen grid diagrams which are congruent with the Archimedean solids.
The inner vertices at the center are hidden. The second row shows the vacuum bubbles for 
each grid diagrams, and the third row shows the 2-wall charge densities ${\cal W}_2$. The
bubbles are dual to the grid diagram, so that their shapes are congruent with the thirteen
Catalan solids.}
\label{fig:Catalan_bubble}
\end{center}
\end{figure}

Fig.~\ref{fig:Catalan_bubble} shows the vacuum bubbles and the 2-wall charge densities
for the Archimedean grid diagrams with an additional 
inner vertex at the center. As desired, the bubble shapes are dual to the grid diagrams,
namely we indeed have the Catalan bubbles.
Let us explain how to construct these bubbles by taking the truncated tetrahedron
(the solids at the left-top corners in Fig.~\ref{fig:Catalan_bubble}) as an example.
We need $N = 12+1$ flavors, and put the twelve masses on the twelve vertices of the
truncated tetrahedron. In addition, we put the thirteenth mass at the center of the
truncated tetrahedron. Then we take the moduli matrix $H_0 = (1,1,\cdots,1, e^{a_{13}})$
with $e^{a_{13}}$ being sufficiently large. This way, we get the triakis tetrahedron as
the vacuum bubble, dual to the truncated tetrahedron.
All the rest solids can be obtained by similar procedures. 

\begin{figure}[t]
\begin{center}
\includegraphics[width=15cm]{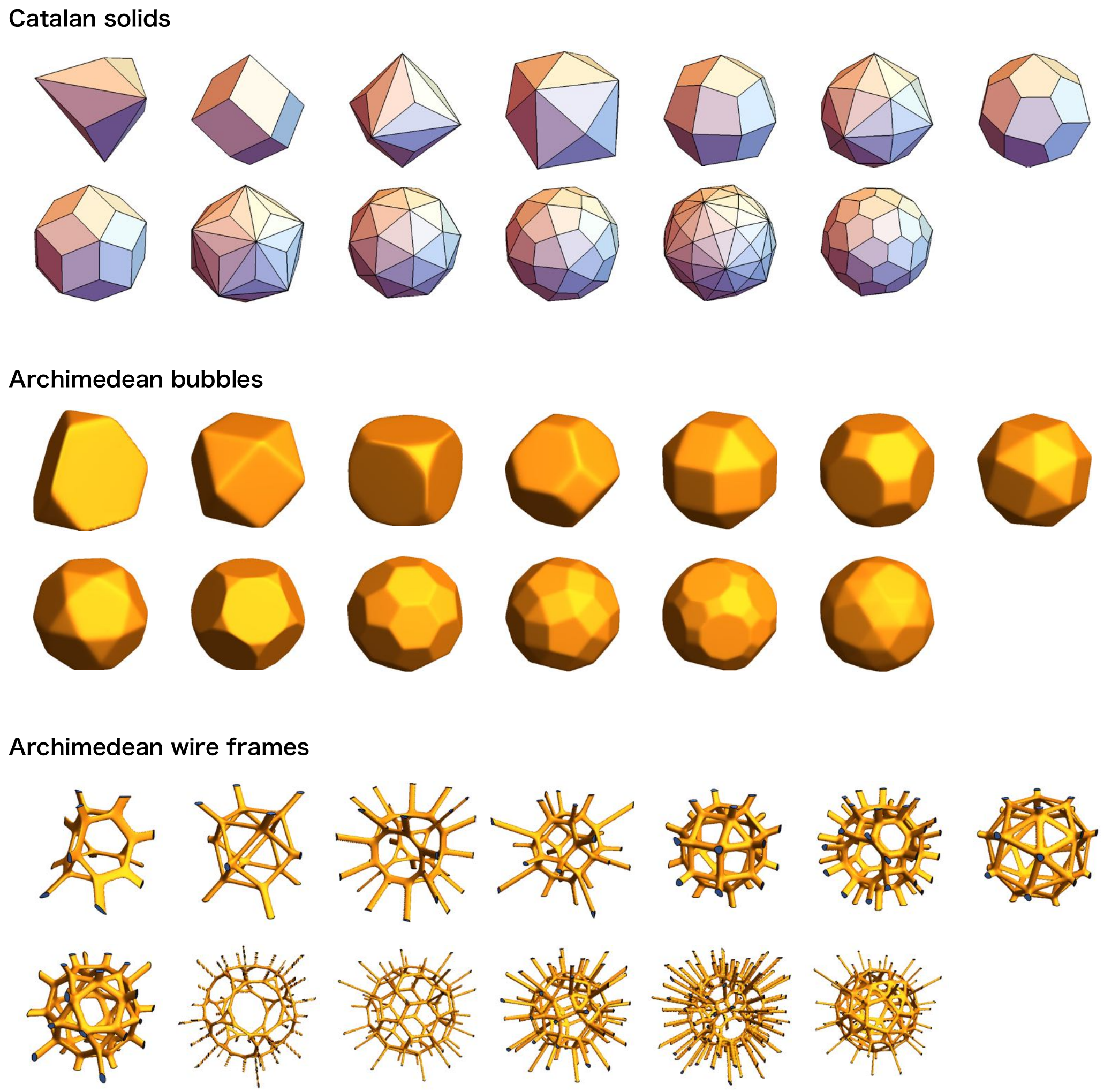}
\caption{The top row shows thirteen grid diagrams which are congruent with the Catalan solids.
The inner vertices at the center are hidden. The second row shows the vacuum bubbles for 
each grid diagrams, and the third row shows the 2-wall charge densities ${\cal W}_2$. The
bubbles are dual to the grid diagram, so that their shapes are congruent with the thirteen
Archimedean solids.}
\label{fig:Archimedean_bubble}
\end{center}
\end{figure}
Fig.~\ref{fig:Archimedean_bubble} shows the vacuum bubbles and the 2-wall charge densities
for the Catalan grid diagrams with an additional 
inner vertex at the center. As expected, the bubble shapes are dual to the grid diagrams,
namely we have the Archimedean bubbles.
Let us explain how to construct these bubbles by taking the triakis tetrahedron
(the solids at the left-top corners in Fig.~\ref{fig:Archimedean_bubble}) as an example.
We need $N = 8+1$ flavors, and put the eight masses on the eight vertices of the
trikias tetrahedron. In addition, we put the ninth mass at the center of the
trikias tetrahedron. Then we take the moduli matrix $H_0 = (1,1,\cdots,1, e^{a_{9}})$
with $e^{a_{9}}$ being sufficiently large. This way, we get the truncated tetrahedron as
the vacuum bubble, dual to the trikias tetrahedron.
All the rest solids can be obtained by similar procedures.

\subsection{The Kepler-Poinsot vacuum bubbles}
\begin{figure}[t]
\begin{center}
\includegraphics[width=11cm]{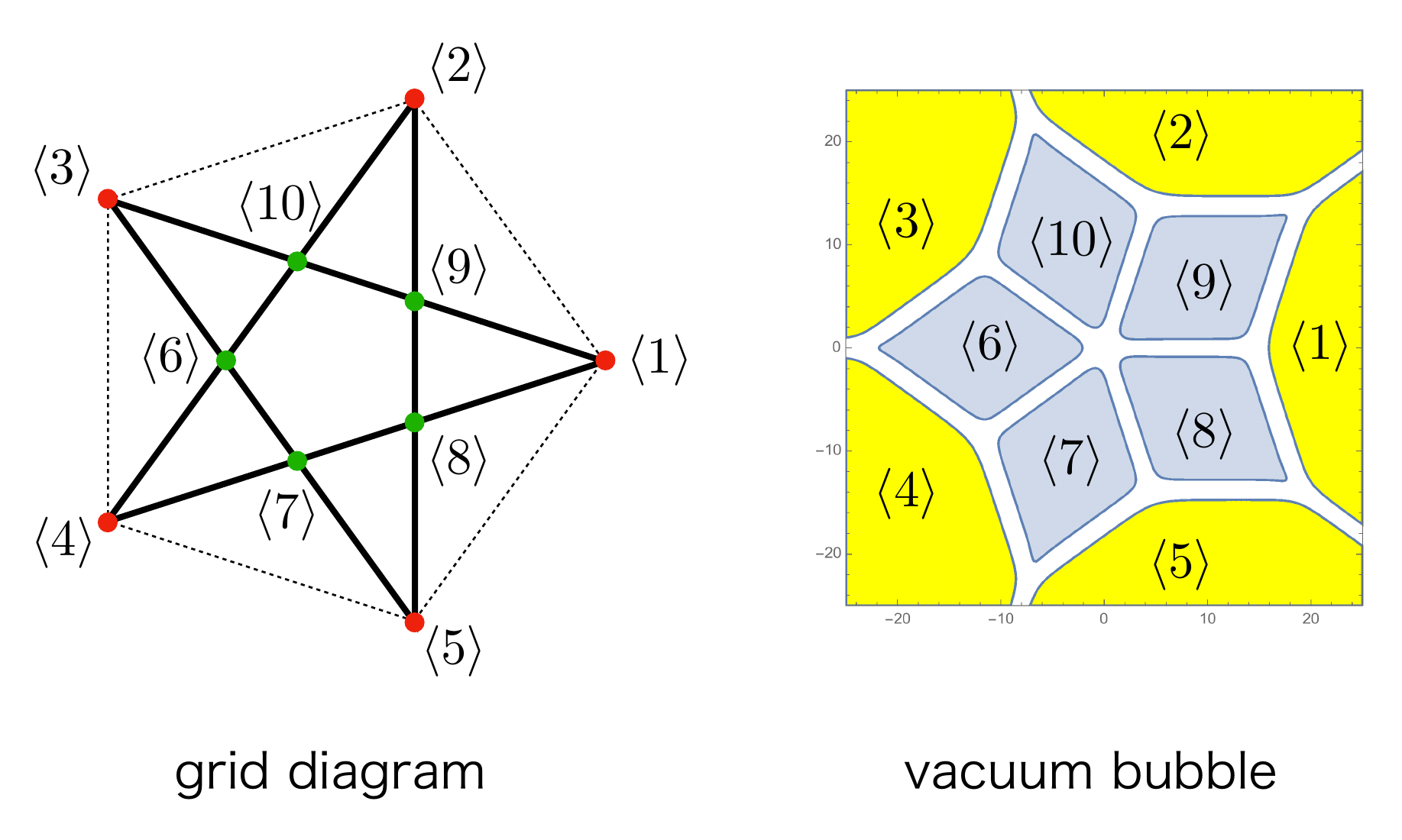}
\caption{The left panel shows the grid diagram congruent with a pentagram. The right panel
shows an exact solution with five vacuum bubbles which have the same influence. The bubbles
form a stellated pentagon.}
\label{fig:pentagram}
\end{center}
\end{figure}
%%%%%%%%%%%%%%%%%%%
\begin{figure}[t]
\begin{center}
\includegraphics[width=12cm]{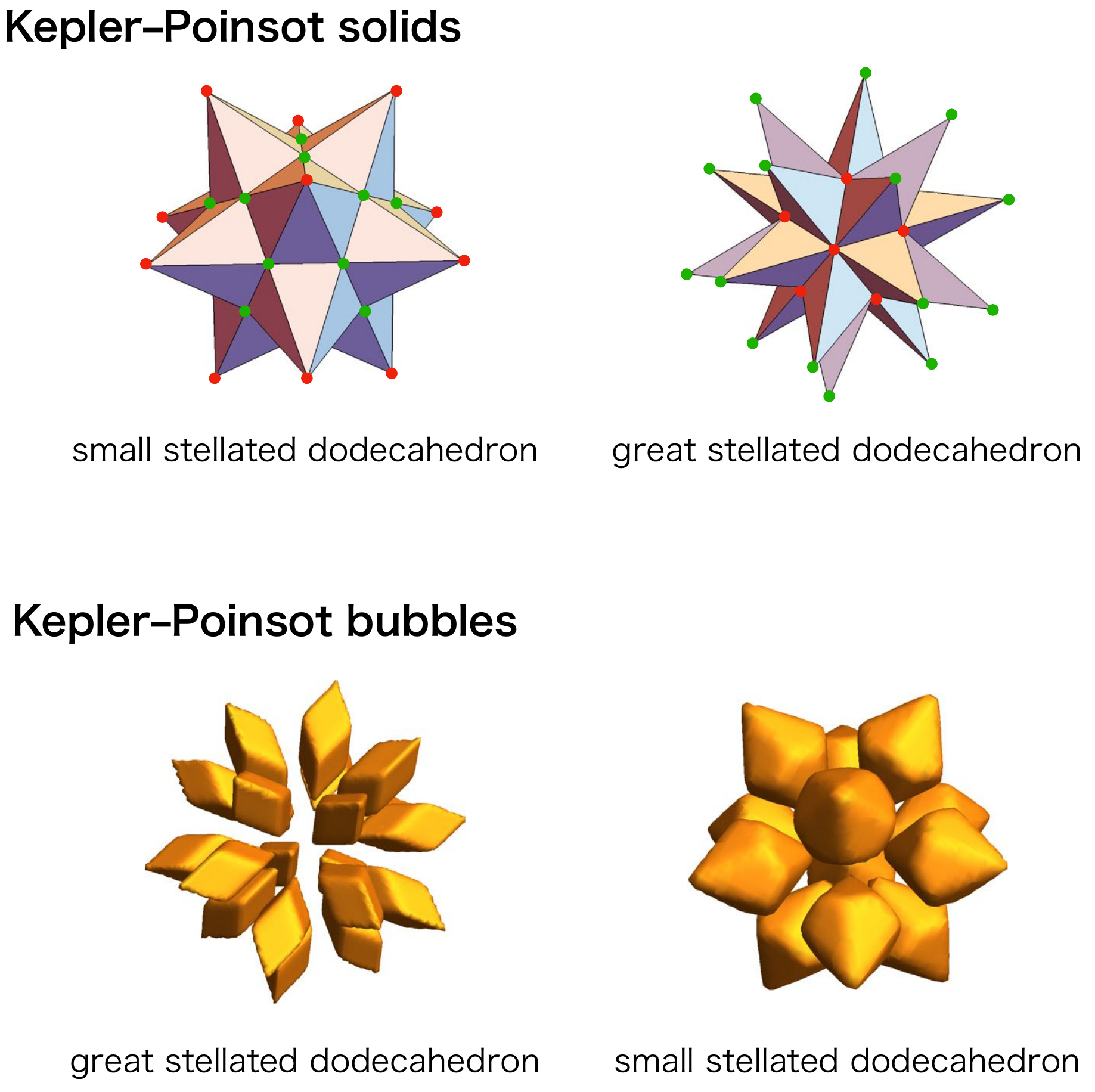}
\caption{The top row: The two regular star solids categorized 
into the Keplar-Poinsot solids are used as the grid diagrams.
The bottom row: The corresponding vacuum bubbles. The small (great) stellated
dodecahedron yields the great (small) stellated dodecahedral bubble.}
\label{fig:kepler_poinsot_solid}
\end{center}
\end{figure}
%%%
So far, we have only studied the convex polyhedra. Before closing this section,
let us briefly mention star polyhedra. Here, we take the small and great stellated dodecahedron.
Namely, we consider the grid diagrams which are congruent with the star polyhedra.
Stellating a dodecahedron to a stellated dodecahedron 
is a three-dimensional analogue of stellating a pentagon to a pentagram.

In order to obtain an intuitive picture, let us first consider two dimensional
domain wall network for a grid diagram which is congruent with a pentagram.
The ten vacua are located on vertices of the pentagram as shown in Fig.~\ref{fig:pentagram}.
The pentagram consists of large (the red five vertices) and small (the green five vertices)
pentagons. The five vacua corresponding to the
small pentagon lead to the five vacuum bubbles. When we equally broaden them by
the moduli matrix $H_0 = (1,1,1,1,1,e^{a},e^{a},e^{a},e^{a},e^{a})$ with sufficiently 
large $e^{a}$, the bubbles ($\left<6\right>$, $\left<7\right>$,
$\left<8\right>$, $\left<9\right>$, $\left<10\right>$) 
form a stellated pentagon as shown in Fig.~\ref{fig:pentagram}.

We are ready to study three dimensional domain wall networks with
the grid diagrams congruent with the small and great stellated dodecahedron
shown in Fig.~\ref{fig:kepler_poinsot_solid}.
The former consists of a large icosahedron with  twelve red vertices and 
a small dodecahedron with twenty green vertices. 
The vertices of the small dodecahedron are inner vacua. Therefore, by broadening
them with the equal weight, we have twenty vacuum bubbles which form the
great stellated dodecahedron as shown in the left-bottom panel of Fig.~\ref{fig:kepler_poinsot_solid}.
On the other hand, the latter is made of
the large dodecahedron with the green vertices and the small icosahedron with
the red vertices. Now, the vertices of the dodecahedron become the inner vacua.
So, when we equally broaden them, we have twelve vacuum bubbles which form the
small stellated dodecahedron as shown in the right-bottom panel of Fig.~\ref{fig:kepler_poinsot_solid}.
In short, when the grid diagram is the small stellated dodecahedron, the vacuum bubbles 
form the great stellated dodecahedron, and vice versa. Although the small and great
stellated dodecahedrons are not dual to each other (dual of the small stellated dodecahedron is
the great dodecahedron while that of the great stellated dodecahedron is the great icosahedron), here, we found
they exchange via the vacuum bubbles.

%%%%%%%%%%%%%%%%% S E C T I O N 7 %%%%%%%%%%%%%%%%%%

\section{Summary and discussion}  
\label{sec:conclusion}

In this paper, we have proceeded study on the non-planar BPS domain wall junctions 
in the Abelian gauge theory with
the $N $ Higgs fields $H_A$ and the $D$ neutral scalar fields $\Sigma_m$ in $D+1$ dimensions, 
which were recently proposed in Ref.~\cite{Eto:2020vjm}.
In the previous work \cite{Eto:2020vjm}, we obtained 
the new exact BPS solutions of the {\it single}
domain wall junctions associated 
with a particular symmetry breaking pattern ${\cal S}_{D+1} \to {\cal S}_D$ 
(${\cal S}_D$ is the symmetric group of rank $D$) 
in $D+1$ dimensions. 
We also needed to impose the symmetry ${\cal S}_{D+1}$ in addition to the special relation 
between the model parameters (the masses, the gauge coupling, and the Fayet-Illiopoulos term) for having the exact solutions in \cite{Eto:2020vjm}. 

Generalizing the previous study, 
the present work has 
dealt with the {\it generic network of domain walls including multiple junctions}. 
We have not imposed any particular discrete symmetry in this paper. 
We have succeeded in solving partially the BPS equations by the moduli matrix formalism
and finding all the moduli parameters of the generic domain wall network solutions.
While the master equation (\ref{eq:master}) cannot be analytically solved in general,
we focused on the infinite $U(1)$ gauge coupling limit
in which the model reduces to 
the $\mathbb{C}P^{N -1}$ model  
and the BPS equations including the master equation are 
fully solvable for any moduli parameters.
These are the first analytic solutions for non-planar
domain wall networks in $D$ dimensions ($D\ge 3$).

As demonstration on showing how the non-planar networks look like, 
we have studied the $\mathbb{C}P^{N -1}$
model in $D=3$ for $N =4,5,6$ in details.
In the case of $N  = 4$, the solution has only one junction. The solution is similar to
one found in \cite{Eto:2020vjm}, but is more generic. The solutions
in this paper can be obtained for any grid diagrams congruent with tetrahedra in contrast to the previous case \cite{Eto:2020vjm} in which 
the grid diagram was limited to the regular tetrahedron.
The network structure appears for $N  > 4$. For $N  = 5$, we have found two different types of networks in general. The first type has a vacuum bubble (a compact domain)
surrounded by the four semi-infinite  vacuum domains. The other type does not have any
bubbles but all the five vacuum domains are semi-infinitely extended.
The corresponding grid diagram is the tetrahedron with an inner vertex for the former
and the dipyramid for the latter. We have shown the network shape is controlled by  
one moduli parameter. 
We also have studied 
the special cases in which the four vertices are on a plane while the fifth vertex is 
off the plane. 
In the $N =6$ case, there are three different types in general.
With respect to the grid diagram, they correspond to the tetrahedron with two inner
vertices, the dipyramid with one inner vertex, and the octahedron without any inner vertices.
Accordingly, there appear two, one, and no vacuum bubbles in the networks, respectively.
As in the case of $N =5$, we have explicitly shown how the network shape changes according with the moduli parameters. They are essentially controlled by the two moduli besides the translations, either size of bubble or distance between junctions.

Finally, we have constructed the beautiful polyhedra known from ancient times
as domain wall networks. We have started with the grid diagrams congruent with
the five Platonic solids with an inner vertex at the centers.
The network solutions have the single vacuum bubbles corresponding to the inner vertices,
and we have found that shapes of the bubble are dual to the grid diagrams.
In addition to the regular polyhedra, we have also investigated the semi-regular polyhedra,
the Archimedean solids. The corresponding vacuum bubbles are again dual to the
Archimedean, namely, the Catalan solids. 
Conversely,
the grid diagrams congruent with the Catalan solids lead to the bubbles congruent with the Archimedean solids.
Our final examples have been star polyhedra. We have taken two well-known star polyhedra from
the Kepler-Poinsot solids, the small and great stellated dodecahedrons.
The former (latter) has twenty (twelve) inner vertices.
Accordingly, the same number of the bubbles appears in the networks.
Interestingly, the bubbles in the former case form the great stellated dodecahedron 
and those in the latter case form the small stellated dodecahedron.

Before closing this work, 
let us make several comments on future directions.
First, although we have established the generic formulae for the exact solutions of
the domain wall networks in the limit of the $\mathbb{C}P^{N -1}$ model
in generic $D+1$ dimensions, we only have shown concrete configurations in the $D=3$ case.
For $D \ge 4$, the networks become more complicated and their deformations by
changing moduli parameters are intricate. We will explain such higher dimensional
domain wall networks in more details elsewhere.

Second, we only have studied the Abelian gauge theories and the massive $\mathbb{C}P^{N -1}$ 
model as the infinite  gauge coupling limit in this paper. The non-Abelian generalization
was obtained in $D=2$ \cite{Eto:2005cp,Eto:2005fm}. We will study non-Abelian non-planar
domain wall networks in higher dimensions $D\ge3$ elsewhere.

Third, we have met the polyhedra 
and well-known mathematical notions like the duality in study of the domain walls.
While the mathematical solids are sharp objects, the vacuum bubbles found in this 
work are rounded off. We expect that the domain wall networks would mathematically useful for studying such melting polyhedra and polytopes
as the case of Amoeba and tropical geometry discussed in $D=2$ \cite{Fujimori:2008ee}.

Fourth, the grid diagrams in the $\bm{\Sigma}$ space and the networks in $\bm{x}$ space
seem to be very similar to the Delaunay diagram and the Voronoi diagram appearing in vast
area of Science. 

We expect our exact solutions of the $D$ dimensional domain wall networks and the grid diagrams would be applicable to many areas.
Josephson junctions of superconductors
\cite{Nitta:2012xq,Fujimori:2016tmw} are one of interesting applications.

%%%%%%%%%%   ACKNOWLEDGMENTS   %%%%%%%%%%

\section*{Acknowledgements}

This work was supported by the 
Ministry of Education, Culture, 
Sports, Science (MEXT)-Supported Program for the Strategic Research 
Foundation at Private Universities ``Topological  Science'' 
(Grant  No.~S1511006) 
and JSPS KAKENHI Grant Numbers 16H03984.
The work of M.~E.~ is also supported in part 
by JSPS Grant-in-Aid for Scientific Research 
KAKENHI Grant No. JP19K03839, and
by MEXT KAKENHI Grant-in-Aid for 
Scientific Research on Innovative Areas
``Discrete Geometric Analysis for Materials Design'' No. JP17H06462 from the MEXT of Japan.
The work of M.~N.~is also supported 
in part by JSPS KAKENHI Grant Number 18H01217 
by a Grant-in-Aid for Scientific Research on Innovative Areas 
``Topological Materials Science'' 
(KAKENHI Grant No.~15H05855) from MEXT of Japan.

%%%%%%%%%%%%%%%%% A P P E N D I X %%%%%%%%%%%%%%%%%%
%\begin{appendix}
%
%\end{appendix}

\bibliographystyle{jhep}

%\bibliography{references}
\end{document}